\documentclass[pdflatex,sn-mathphys-num]{sn-jnl}

\usepackage{graphicx}%
\usepackage{multirow}%
\usepackage{amsmath,amssymb,amsfonts}%
\usepackage{amsthm}%
\usepackage{mathrsfs}%
\usepackage[title]{appendix}%
\usepackage{xcolor}%
\usepackage{textcomp}%
\usepackage{manyfoot}%
\usepackage{booktabs}%
\usepackage{algorithm}%
\usepackage{algorithmicx}%
\usepackage{algpseudocode}%
\usepackage{array}
\usepackage{listings}%
\usepackage{longtable}
\usepackage{makecell}
\usepackage{hhline}
\usepackage{tikz}
\usepackage[normalem]{ulem}
\usetikzlibrary{positioning}
\usepackage{float}
\theoremstyle{thmstyleone}%
\theoremstyle{thmstyletwo}%

\theoremstyle{thmstylethree}%

\newcounter{ToDo}
\newcounter{gaocomm}
\newcounter{wangcomm}
\newcounter{Note}
\definecolor{blue-violet}{rgb}{0.00,0.75,0.90}
\definecolor{mygreen}{rgb}{0.0, 0.5, 0.0}
\definecolor{awesome}{rgb}{1.0, 0.13, 0.32}
\definecolor{bostonuniversityred}{rgb}{1.0, 0.0, 0.0}

\begin{document}

\title[Carbon Credits Engineering]{Engineering Carbon Credits with AI Towards A Responsible FinTech Era: The Practices, Implications, and Future}

\author[1]{\fnm{Qingwen} \sur{Zeng}}\email{qzen5227@uni.sydney.edu.au}\equalcont{These authors contributed equally to this work.}
\author[1]{\fnm{Hanlin} \sur{Xu}}\email{haxu3874@uni.sydney.edu.au}\equalcont{These authors contributed equally to this work.}
\author[2]{\fnm{Nanjun} \sur{Xu}}\email{nanjun.xu@adelaide.edu.au}
\author[1]{\fnm{Zhenghao} \sur{Zhao}}\email{zzha0595@uni.sydney.edu.au}
\author[4]{\fnm{Joakim} \sur{Westerholm}}\email{joakim.westerholm@sydney.edu.au}
\author[3]{\fnm{Flora} \sur{Salim}}\email{flora.salim@unsw.edu.au}
\author[4]{\fnm{Junbin} \sur{Gao}}\email{junbin.gao@sydney.edu.au}
\author[1]{\fnm{Huaming} \sur{Chen}}\email{huaming.chen@sydney.edu.au}

\affil[1]{\orgdiv{School of Electrical and Computer Engineering}
          \orgname{The University of Sydney}
          \city{Camperdown}\postcode{NSW 2006}
          \country{Australia}}
          
\affil[2]{\orgdiv{Adelaide Business School}
          \orgname{The University of Adelaide}
          \city{Adelaide}\postcode{SA 5005}
          \country{Australia}}
          
\affil[3]{\orgdiv{School of Computer Science and Engineering}
          \orgname{University of New South Wales Sydney}
          \city{Kensington}\postcode{NSW 2033}
          \country{Australia}}
          
\affil[4]{\orgdiv{The University of Sydney Business School}
          \orgname{The University of Sydney}
          \city{Camperdown}\postcode{NSW 2006}
          \country{Australia}}

\abstract{
Carbon emissions drive climate change, and carbon credits mitigate climate deterioration and environmental damage while assisting organizations in managing their carbon footprint. Fully utilizing carbon credits remains challenging. This study enhances understanding of the engineering practices for carbon credits to develop responsible fintech solutions and provide insights for carbon emission management. We review the negative impacts of organizations' strategy of evading carbon management through non-disclosure of carbon emissions. Evidence shows that both non-disclosure of carbon emissions and high carbon emissions negatively affect an organization's financial stability and market value, suggesting that organizations should manage carbon emissions and transparently share information to mitigate risks. We examine engineering methods for more cost-effective carbon management, focusing on data-driven computing solutions: factors influencing carbon prices, carbon price prediction algorithms for optimized carbon credit purchasing strategies, and corporate carbon emission prediction algorithms. These methods enable performance assessments for investors and governments when carbon emissions data is not disclosed and help organizations estimate future carbon credits needs for budget planning and strategy optimization. Finally, integrating carbon price and carbon emission predictions, we propose future research directions, including prediction of corporate-level carbon management costs, laying a foundation for quantitative research on how carbon management practices impact corporate market value and financial performance. This systematic review provides a comprehensive synthesis of carbon credits with a unique focus on computing solutions and engineering practices, highlighting AI's role in enhancing transparency and fostering social accountability for an inclusive and trustworthy low-carbon transition.
}


\maketitle

\section{Introduction}

In today's global landscape, the Industrial Revolution marked a turning point where fossil fuels became the primary energy source, driving economic growth but also significantly increasing carbon emissions \cite{carbon1}. This rise in carbon emissions has led to serious climate issues, as excessive carbon dioxide forms a layer around the Earth's surface. This layer traps more heat while reducing its dissipation, directly leading to the global temperature increases. The resulting climate change is profound and often irreversible effects on ecosystems, including glacier melting and biodiversity loss. Additionally, rising sea levels also pose a grave threat to human settlements, particularly in coastal regions. These environmental shifts are not merely technical or economic challenges, but profound social crises that necessitate a re-evaluation of the ethical contract between technology, finance, and society. In response, the global community has invested substantial resources in promoting climate-friendly solutions \cite{carbon2}. A key initiative, the Paris Agreement, reflects the collective commitment of nations to limit global temperature increases to within 2 degrees Celsius by the end of this century \cite{carbon3}.

The Carbon Emissions Trading (CET) system, proposed under the Kyoto Protocol, is a key strategy to combat climate change. Introduced in 1997, CET aims to reduce greenhouse gas emissions by implementing a `quota trading' mechanism. Each unit of carbon credit allows the emission of one ton of carbon dioxide \cite{carbon4}. Specific emission quotas are allocated to countries based on their respective development status and goals, which is then distributed to domestic enterprises and organizations. Entities that exceed their allocated emissions must purchase carbon credits on the carbon trading market, while those with surplus credits can sell them on the market. This scheme has been adopted in many regions, including Europe, Asia, and the Americas \cite{carbon5}. In addition to regulating global carbon emissions, CET incentivizes innovation in green technologies and promotes the use of renewable energy sources to further reduce carbon emissions.

As the carbon credit mechanism matures, its role in mitigating global temperature rise is increasingly recognized. Decision-making must now integrate strategies for managing both carbon emissions and the carbon credits cost. For organizations with high emissions, the cost of purchasing additional carbon credits can be substantial \cite{carbonmanagement}, making accurate carbon credit price prediction essential for effective cost management \cite{carbonprice}.

Additionally, many organizations opt to either inaccurately disclose their carbon emissions or avoid voluntarily disclosing their actual carbon emissions in practice \cite{carbon69}. This creates significant risks for businesses, particularly when managing carbon emission responsibilities that are closely tied to corporate reputation and financial performance. For example, some organizations fail to actively manage their carbon emissions, particularly by not purchasing carbon credits to offset excess emissions. Such practices not only clearly violate policy regulations but also expose organizations to greater compliance risks. To avoid the potential damage to their corporate image, such as fines, a decrease in market value, and harm to their reputation, many organizations choose not to disclose their actual carbon emissions, carbon management practices, or the associated operational costs in their financial reports \cite{carbon72}. In this paper, our findings reveal that better engineering the carbon credits for industry sectors can enable an enhanced transparency of the process, which can significantly boost the organization performance.

There are two key reasons for the increased transparency in carbon emissions disclosure. First, in the financial markets, increasing attention is being paid to whether organizations adhere to ESG (Environmental, Social, and Governance) principles. Investors, particularly institutional investors, are placing greater emphasis on organizations' performance in sustainable development \cite{carbon68}. They are more inclined to allocate funds to organizations that can create long-term social value and mitigate negative externalities. This trend has significantly heightened the demand for ESG information disclosure, with investors favoring organizations that actively implement ESG principles. Consequently, financial investors typically rely on the information disclosed in organizations' financial reports to assess their ESG performance. However, some organizations have not fully disclosed their carbon emissions and carbon management practices, making it difficult for investors to gain a comprehensive understanding of the company's actual operations and true value. This information asymmetry not only increases uncertainty and risk in investment decisions but also potentially leads investors to miss identifying high-risk organizations. Therefore, having accurate and comprehensive information on a company's carbon management is crucial for investors to make informed decisions. Furthermore, the integration of AI into carbon credit engineering transcends corporate profit-seeking; it empowers public oversight and reinforces social trust by mitigating the risks of greenwashing. This alignment of data-driven precision with collective environmental welfare is fundamental to the emergence of a truly responsible FinTech era.

Second, government has implemented stricter regulations on organizations' disclosure of carbon emissions. Governments also need to understand a company’s carbon emissions and management practices to formulate more effective carbon management policies and enforce penalties on non-compliant organizations, thereby supporting the achievement of emission reduction targets and controlling global temperature rise. By obtaining accurate carbon emission data, governments can better evaluate the effectiveness of current policies and adjust them in a timely manner to ensure that carbon emissions align with expected targets \cite{carbon70}. Additionally, understanding a company’s carbon management strategies helps governments identify industries or organizations that are lagging in emission reduction efforts, enabling the formulation of targeted support or punitive measures to promote the overall industry's green transformation. Furthermore, governments can use this data in international climate negotiations to showcase the country's efforts and achievements in emission reductions, thereby strengthen international confidence in the country's climate commitments. 

\begin{figure}[htbp]
    \centering
    \includegraphics[width=0.85\textwidth]{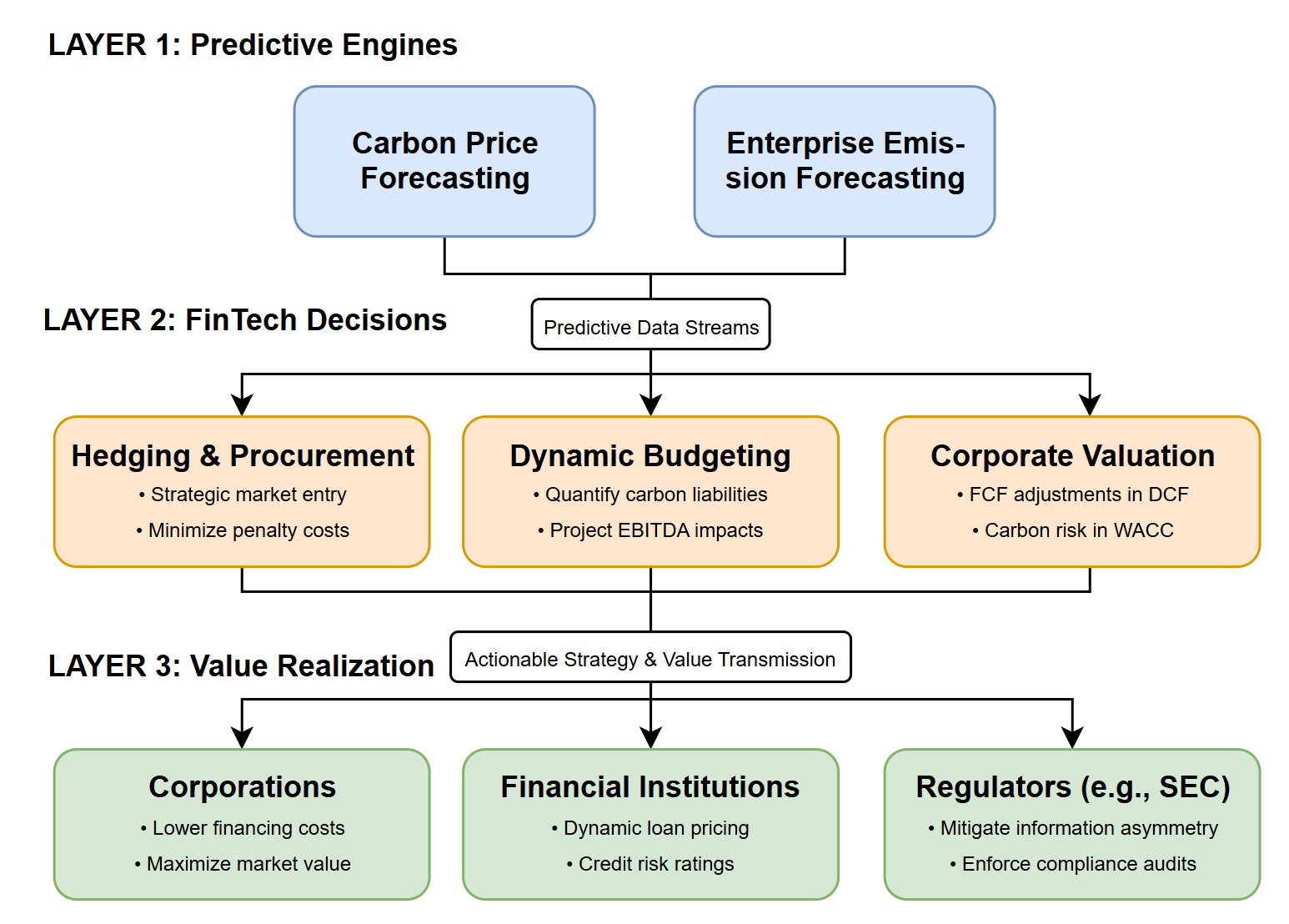}
    \caption{The unified conceptual framework bridging data inputs, algorithmic predictions, and FinTech applications in the carbon market.}
    \label{fig:framework}
\end{figure}
To bridge the conceptual gap between technical prediction algorithms and corporate financial strategy, this review proposes a unified framework tailored for the ``Responsible FinTech Era,'' as illustrated in Fig.~\ref{fig:framework}. Instead of viewing emission disclosures, market drivers, and forecasting models in isolation, our review maps them across a continuous operational pipeline: (1) \textbf{Layer 1: Predictive Engines,} which process market data and macroeconomic indicators (explored in Section 4) through advanced algorithms to generate precise carbon price and enterprise emission forecasts (Sections 5 and 6); (2) \textbf{Layer 2: FinTech Decisions,} which translates these predictive outputs into actionable corporate strategies, such as procurement timing, dynamic compliance budgeting, and Discounted Cash Flow (DCF) valuation (discussed in Section 7); and (3) \textbf{Layer 3: Value Realization,} which demonstrates how these financial decisions ultimately create tangible benefits for key stakeholders—lowering financing costs for corporations, enabling dynamic risk ratings for financial institutions, and enhancing market transparency for regulators (detailed in Section 3). By explicitly mapping predictive precision to financial use-cases and stakeholder value, this framework addresses the complex challenges faced by organizations, investors, and governments. Guided by this pipeline, we specifically investigate the following three core areas:

\begin{enumerate}
    \item \textbf{The impact of non-disclosure of carbon emissions and carbon management information on organizations. This study explores whether non-disclosure can effectively help organizations avoid penalties and negative consequences when they fail to manage carbon emissions as required.}
    \begin{itemize}
        \item \textbf{Findings:} This study finds that organizations that fail to disclose information about their carbon emissions and management not only risk negative impacts but may also face even more severe consequences. Therefore, evasion and non-disclosure are unsustainable strategies. Instead, organizations need to proactively manage carbon emissions, fulfill their social responsibilities, and transparently disclose relevant information. This approach not only helps organizations enhance regulatory compliance and market trust, but also allows them to gain a competitive advantage in markets that increasingly prioritize ESG principles.
    \end{itemize}
    
    \item \textbf{The latest carbon credit price prediction algorithms and factors influencing carbon prices.}
    \begin{itemize}
        \item \textbf{Findings:} This study systematically reviews and analyzes the state-of-the-art methods for predicting carbon credit prices. These methods help organizations accurately identify optimal times to purchase carbon credits when prices are low, thereby reducing carbon management costs and providing economic incentives for optimizing carbon management. By lowering the cost of purchasing carbon credits, organizations can reduce expenditures while boosting profits, which in turn encourages them to voluntarily and transparently disclose their carbon emissions. This process not only helps organizations improve their compliance and social responsibility but also strengthens the trust of investors and regulators in the company’s carbon management efforts, fostering a virtuous cycle of proactive corporate participation in the carbon market.
    \end{itemize}
    
    \item \textbf{The latest corporate carbon emission prediction algorithms.}
    \begin{itemize}
        \item \textbf{Findings:} This study systematically reviews the most effective corporate carbon emission prediction methods available. These methods provide accurate emission assessments even when organizations fail to disclose or misreport carbon emission data. Through these prediction methods, financial investors can more precisely evaluate an organization's environmental performance and long-term value, leading to more informed investment decisions. Simultaneously, governments can use these data to develop more effective carbon emission policies, ensuring that regulatory measures are implemented and enforced. Additionally, these methods enable governments to identify and monitor an organization's actual carbon emissions, allowing for appropriate management and penalties when necessary. This approach not only reinforces corporate accountability but also enhances market transparency, ultimately contributing to the achievement of global emission reduction targets.
    \end{itemize}
\end{enumerate}

Although the existing literature increasingly focuses on the application of carbon credits, there is a lack of comprehensive reviews on this topic. Existing reviews tend to be limited on a specific area \cite{carbon2, carbon74, carbon75, carbon76}, such as the impact of carbon credits on farmers' income in agricultural, or the role of blockchain for carbon credits in the construction industry. However, these studies do not provide a systematic review of carbon credits from a financial and fintech perspective, leaving the broader implication for enterprises, governments, and the financial industry largely unexplored. In particular, there is a significant gap in understanding and leveraging carbon credits in fintech with advanced computing solutions.

This study addresses the existing gap by exploring carbon emission and carbon price prediction technologies, providing stakeholders with more comprehensive management tools. Following the findings of carbon credit price and carbon emission prediction algorithms, we present a thorough understanding of the significance of carbon credits, in particular for the organization performance. We also identify a promising yet challenging task as the future research direction to combine these two important tasks to further enhance the carbon management practices, which will in turn increase the corresponding market value.  

Thus, in this work, we conduct a systematic literature review \cite{carbon10}, to comprehensively analyze the application of carbon credits in fintech sector, providing valuable insights for stakeholders in tackling climate change challenges with state-of-the-art computing algorithms and solutions. The main contributions of this article are as follows:
\begin{enumerate}
    \item This study systematically summarizes the impact of transparency in carbon emission management on corporate development. It explores the strategies that organizations use to mitigate compliance risks by not disclosing carbon emission information, analysing the effectiveness and potential consequences of such approaches. The research shows that a lack of transparency not only fails to protect organizations from negative impacts of policies and markets but may also exacerbate the risks they face. Following sections highlight the current best practices of distilling carbon emission management in a measurable approach, in which we identify the topics of estimations of carbon credit price and corporate carbon emission.
    \item This study provides a comprehensive review of the factors influencing carbon credit prices, offering insights that can qualitatively predict price trends. This contribution aids organizations in better managing their carbon credit purchasing strategies, thereby reducing costs. Additionally, these factors can serve as input variables for algorithms predicting carbon price fluctuations, further enhancing the accuracy and applicability of these algorithms in supporting corporate decision-making in the carbon market.
    \item The article systematically summarizes and critically evaluates the applicability, limitations, and advantages of current carbon credit price prediction algorithms and methods. This analysis supports organizations in making informed decisions about the timing and scale of carbon credit purchases, thereby promoting more effective carbon management practices.
    \item The article reviews and discusses the latest corporate carbon emission prediction algorithms, highlighting their strengths and weaknesses. This contribution assists investors and regulators in accurately assessing a company's environmental performance, even in cases of non-disclosure, thereby reducing information asymmetry and enhancing market transparency.
    \item This study, through an in-depth exploration of carbon emission and carbon price prediction technologies, reveals the significant impact of carbon credits on corporate market value. These technologies not only help organizations strategically plan their carbon credit purchases and reduce carbon management costs, but more importantly, they provide a foundation for quantitative research, enabling organizations to predict the impact of their carbon management practices on financial and market performance. By actively managing carbon emissions and transparently disclosing related information, organizations can enhance market trust, reduce financing costs, and ultimately increase their market value. This lays a strong foundation for future research to quantify the long-term impact of carbon management on corporate market value.
\end{enumerate}

The remaining structure of the article is as follows: Section 2 provides a detailed description of the procedure used to select papers for this systematic review, including the inclusion and exclusion criteria for literature retrieval. Section 3 discusses the impact of corporate transparency in carbon emission disclosure. Section 4 conducts an in-depth analysis of the key factors influencing the price of carbon credits. Section 5 examines algorithms related to predicting carbon credit prices, focusing on their applicability and effectiveness. Section 6 explores methods for predicting corporate carbon emissions, discussing the strengths and weaknesses of various algorithms. Finally, Section 7 synthesizes these findings, comparing this study with relevant reviews and providing a comprehensive discussion on the future prospects of carbon credits in the financial and FinTech fields before concluding the paper.

\section{Research methodology}

\subsection{Research questions}

This study adopts a systematic literature review approach \cite{SLR} to comprehensively identify and analyze the key applications of carbon credits in the Fintech field, as well as the research progress in related areas. To ensure the transparency and replicability of this work, we strictly followed the steps, including initial literature retrieval, setting up inclusion and exclusion criteria, performing study selection process and finally conducting quality assessment. We consider the following research questions as an outline to initiate our investigation:

\begin{enumerate}
    \item What is the impact of corporate carbon emissions disclosure (or non-disclosure) on organizations and society, and what factors influence corporate decisions regarding such disclosures? (Section.~\ref{chp:impacts})

    \item What are the key factors that influence carbon credit prices? (Section.~\ref{chp:factors})

    \item What methods are available for predicting carbon prices? (Section.~\ref{chp:algorithmsCC})

    \item What methods are available for forecasting corporate carbon emissions? (Section.~\ref{chp:algorithmCE})
\end{enumerate}

\subsection{Methodology}

We implemented this work as a systematic literature review following the Preferred Reporting Items for Systematic Reviews and Meta-Analyses (PRISMA) guidelines to improve transparency and reproducibility. The database search was designed to identify primary studies related to carbon-credit pricing, carbon-price prediction, emission-trading markets, corporate carbon-emission disclosure, enterprise-level carbon-emission prediction, and AI/ML-based carbon-related forecasting. Records were retrieved from \textit{Google Scholar}, \textit{Scopus}, \textit{IEEE Xplore}, and \textit{ScienceDirect}. The search covered studies published from 2006 to 2025 and was finalized in December 2025.

Database-specific query strings were constructed for the primary-study search. \textit{Scopus} searches were conducted using the \texttt{TITLE-ABS-KEY} field with Boolean operators and wildcard truncation. \textit{IEEE Xplore} searches were conducted using \texttt{All Metadata} command search with Boolean operators and phrase matching. \textit{ScienceDirect} searches were conducted using exact keywords and phrase searches without wildcard truncation. \textit{Google Scholar} searches were conducted using exact phrases and simple Boolean combinations. Retrieved records were screened for relevance based on title, abstract, keywords, and available metadata. Table~\ref{tab:database-search} summarizes the database-specific search strategies, search fields, publication-year limits, language restrictions, document-type filters, and final search update.

The primary studies were selected according to their direct relevance to the review scope. A study was treated as a primary study if it substantially examined at least one of the following topics: carbon-price or carbon-credit prediction, carbon trading or emission-trading price dynamics, corporate carbon-emission disclosure and its financial implications, enterprise-level carbon-emission prediction, or AI/ML-based modeling for carbon-related financial or managerial analysis. Studies were excluded from the primary evidence base if they were unrelated to carbon markets or corporate carbon-emission analysis, did not involve prediction, forecasting, modeling, or systematic analysis relevant to the review topic, or only provided general policy, technical, or methodological background without contributing direct evidence to the thematic synthesis.

In addition to the database search, supplementary references were included to support parts of the review that were not intended to define the primary evidence base. These references were added through backward citation tracing, forward citation checking, and manual supplementation when they were needed to clarify concepts, explain technical methods, or provide necessary background for interpreting the reviewed studies. The supplementary references mainly covered four types of material. First, methodological sources were used to support the design and reporting of the systematic literature review. Second, foundational algorithmic papers were included when the reviewed studies relied on methods such as support vector machines, random forests, long short-term memory networks, empirical mode decomposition, wavelet-based decomposition, gradient boosting, and optimization algorithms. Third, policy and market-background sources were used to explain carbon-credit systems, emission-trading schemes, carbon-pricing mechanisms, and corporate carbon-disclosure practices. Fourth, closely related review articles were used to position the contribution of this review against existing literature.

\begin{footnotesize}
\setlength{\tabcolsep}{4pt}
\renewcommand{\arraystretch}{0.95}

\begin{longtable}{@{}>{\raggedright\arraybackslash}p{0.16\textwidth}
                  >{\raggedright\arraybackslash}p{0.46\textwidth}
                  >{\raggedright\arraybackslash}p{0.15\textwidth}
                  >{\raggedright\arraybackslash}p{0.18\textwidth}@{}}
\caption{Database-specific search strategy for identifying primary studies (2006--2025)}
\label{tab:database-search}\\
\toprule
\textbf{Database} & \textbf{Search strategy / query strings} & \textbf{Search field} & \textbf{Filters / final search update} \\
\midrule
\endfirsthead

\toprule
\textbf{Database} & \textbf{Search strategy / query strings} & \textbf{Search field} & \textbf{Filters / final search update} \\
\midrule
\endhead

\midrule
\multicolumn{4}{r}{\textit{(Continued on next page)}}\\
\midrule
\endfoot

\bottomrule
\endlastfoot

Scopus &
TITLE-ABS-KEY(("carbon credit*" OR "carbon emission*" OR "carbon allowance*" OR "emission trading" OR "carbon trading" OR "carbon price*") AND ("predict*" OR "forecast*" OR "machine learning" OR "artificial intelligence" OR "AI" OR "deep learning") AND ("price*" OR "financial performance" OR "firm value" OR "carbon disclosure" OR "corporate carbon emission*" OR "enterprise carbon emission*")) &
Title, Abstract, Keywords &
Publication years 2006--2025; English; Articles, Conference Papers, Reviews; 15 December 2025 \\

\midrule

IEEE Xplore &
(("All Metadata":"carbon credit" OR "All Metadata":"carbon credits" OR "All Metadata":"carbon emission" OR "All Metadata":"carbon emissions" OR "All Metadata":"carbon allowance" OR "All Metadata":"carbon allowances" OR "All Metadata":"emission trading" OR "All Metadata":"carbon trading" OR "All Metadata":"carbon price") AND ("All Metadata":"predict*" OR "All Metadata":"forecast*" OR "All Metadata":"machine learning" OR "All Metadata":"artificial intelligence" OR "All Metadata":"deep learning" OR "All Metadata":"AI") AND ("All Metadata":"price" OR "All Metadata":"prices" OR "All Metadata":"financial performance" OR "All Metadata":"firm value" OR "All Metadata":"carbon disclosure" OR "All Metadata":"corporate carbon emission" OR "All Metadata":"enterprise carbon emission")) &
All Metadata / Command Search &
Publication years 2006--2025; English; Journals and Conferences; 15 December 2025 \\

\midrule

ScienceDirect &
("carbon price" OR "carbon trading" OR "emission trading") AND ("prediction" OR "forecasting" OR "machine learning"); 
("carbon credit" OR "carbon credits" OR "carbon allowance") AND ("price" OR "prediction" OR "forecasting"); 
("carbon emission disclosure" OR "corporate carbon disclosure") AND ("financial performance" OR "firm value"); 
("corporate carbon emission" OR "enterprise carbon emission") AND ("prediction" OR "forecasting" OR "machine learning"). &
Title, Abstract, Keywords &
Publication years 2006--2025; English; Research Articles and Review Articles; 15 December 2025 \\

\midrule

Google Scholar &
("carbon price" OR "carbon allowance price" OR "carbon credit price") AND ("prediction" OR "forecasting" OR "machine learning"); 
("carbon emission disclosure" OR "corporate carbon disclosure") AND ("financial performance" OR "firm value"); 
("enterprise carbon emission" OR "corporate carbon emission") AND ("prediction" OR "forecasting" OR "machine learning"); 
and "emission trading carbon price drivers". &
Full text / metadata as indexed by Google Scholar &
Publication years 2006--2025; English; relevance-based screening; 15 December 2025 \\

\end{longtable}
\end{footnotesize}

To facilitate this process and minimize selection bias, we established strict, operationalized inclusion and exclusion criteria. For inclusion, we considered peer-reviewed, English-only original research articles, conference papers, and case studies that provided empirical data analysis or novel methodological insights into the carbon credit market, price dynamics, corporate carbon disclosure, carbon price forecasting, or enterprise-level emissions prediction. For exclusion, we first removed all duplicate studies. Crucially, rather than relying on subjective assessments of market maturity, we excluded studies focused on experimental or voluntary local markets that lacked formal regulatory frameworks or verifiable trading volumes (e.g., markets not tracked by institutional registries), ensuring consistency with global trends in compliance carbon market development. In practice, this rule led us to exclude studies that analyzed purely experimental trading platforms, local voluntary offset schemes without institutional registration, simulated carbon markets without observed transaction records, and project-level offset mechanisms whose trading volume or allowance issuance could not be independently verified. These studies were excluded because their pricing behavior could not be reliably compared with regulated or institutionally tracked carbon markets, and their inclusion would have introduced substantial heterogeneity into the synthesis.

\begin{figure}[htbp]
\centering
\includegraphics[width=0.90\textwidth]{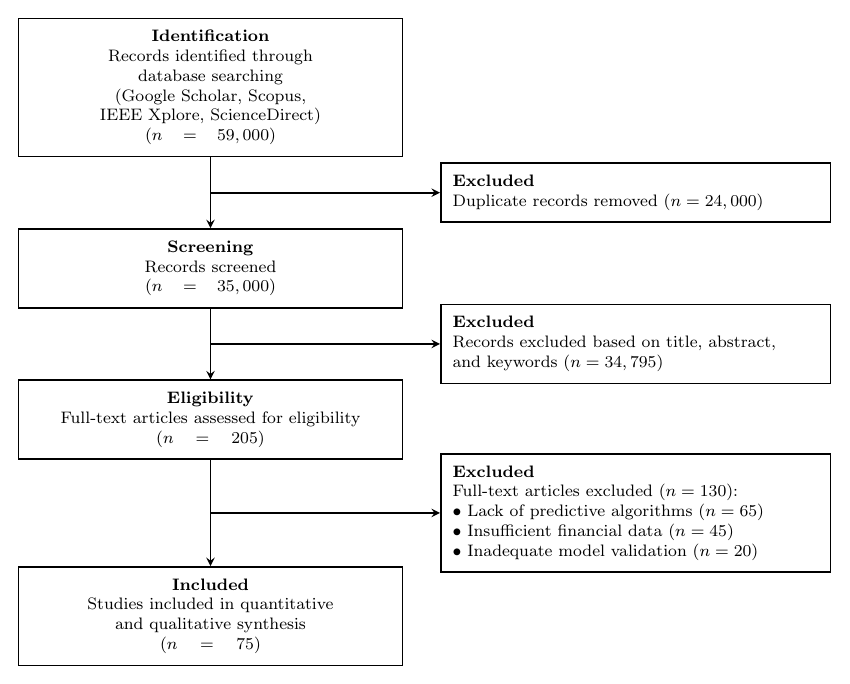}
\caption{PRISMA flow diagram outlining the systematic literature review screening process.}
\label{fig:prisma}
\end{figure}

The screening and selection process, as detailed in the PRISMA flow diagram (see Fig.~\ref{fig:prisma}), yielded an initial total of 59,000 records from the specified databases. After removing approximately 24,000 duplicate records, 35,000 articles remained for the initial screening phase. During this stage, two independent reviewers assessed the titles, abstracts, and keywords, excluding 34,795 papers that were clearly out of scope, which led to the selection of 205 potentially relevant studies. Subsequently, a detailed full-text review was conducted, during which 130 papers were excluded for specific reasons: lack of predictive algorithmic frameworks ($n=65$), insufficient financial or corporate-level data ($n=45$), and inadequate model validation ($n=20$). This process resulted in 75 studies being retained for subsequent analysis.

To ensure the robustness and consistency of the synthesis, the retained studies were further assessed using a standardized quality appraisal rubric. Two independent reviewers evaluated each study across three dimensions: dataset reliability, methodological rigor, and financial or enterprise relevance. Disagreements were resolved through discussion, and a third reviewer was consulted when consensus could not be reached. Each dimension was scored on a 0--3 scale according to the operational criteria summarized in Table~\ref{tab:quality-appraisal-compact}. The maximum quality score for each study was therefore 9. Studies were retained for the main synthesis only when they reached a minimum threshold of 6/9 and did not receive 0 in any single dimension.

\begin{footnotesize}
\setlength{\tabcolsep}{4pt}
\renewcommand{\arraystretch}{0.95}

\begin{longtable}{@{}p{0.22\textwidth}p{0.08\textwidth}p{0.63\textwidth}@{}}
\caption{Detailed 0--3 scoring rubric for quality appraisal (compact layout)}
\label{tab:quality-appraisal-compact} \\
\toprule
\textbf{Dimension} & \textbf{Score} & \textbf{Operational criteria / rationale} \\
\midrule
\endfirsthead

\toprule
\textbf{Dimension} & \textbf{Score} & \textbf{Operational criteria / rationale} \\
\midrule
\endhead

\midrule
\multicolumn{3}{r}{\textit{(Continued on next page)}}\\
\midrule
\endfoot

\bottomrule
\endlastfoot

Dataset Reliability & 0 & Data source is unknown, unverifiable, or absent. \\
                     & 1 & Data source is partially described; coverage, period, or market is limited or unclear. \\
                     & 2 & Data source is reliable and largely transparent; minor gaps or partial preprocessing unclear. \\
                     & 3 & Data source is authoritative and fully documented; coverage, period, frequency, and preprocessing are transparent and reproducible. \\
\midrule

Methodological Rigor & 0 & Model description missing or unclear; no validation or baseline comparison. \\
                     & 1 & Partial model description; validation or baseline is weak or incomplete. \\
                     & 2 & Model fully described; validation and baseline reasonable; some robustness or statistical tests provided. \\
                     & 3 & Model pipeline fully detailed; validation design robust, baseline included, statistical tests and robustness checks present; fully reproducible. \\
\midrule

Financial / Enterprise Relevance & 0 & No link to financial, corporate, or market applications. \\
                                & 1 & Minimal discussion of financial or enterprise implications; not quantified. \\
                                & 2 & Partial quantitative implications; limited market or corporate context. \\
                                & 3 & Clear and quantified implications; broadly applicable to corporate decision-making, FinTech, or market strategies. \\
\end{longtable}
\end{footnotesize}

The quality appraisal was used as a structured screening and interpretation framework rather than as a numerical meta-analytic weighting scheme. Its main purpose was to ensure that all studies retained in the final synthesis met a minimum level of empirical transparency, methodological adequacy, and financial or enterprise relevance. After appraisal, the quality scores of the 75 retained studies showed a generally robust distribution: 10 studies received the maximum score of 9, 25 studies scored 8, 26 studies scored 7, and 14 studies met the baseline threshold with a score of 6. The mean quality score was 7.41 out of 9, indicating that most retained studies exceeded the minimum inclusion threshold while still preserving methodological diversity across the review corpus. The same appraisal dimensions also informed the later critical discussion by drawing attention to recurring methodological issues, such as unclear data provenance, limited validation design, insufficient baseline comparison, and weak connection to practical FinTech or enterprise decision-making. Therefore, the appraisal process supported a more consistent interpretation of the included studies without assigning explicit quantitative weights to individual findings. This appraisal process ensured that the final synthesis was based on studies that were both thematically relevant and methodologically defensible. In the end, we identified 75 studies that met the predefined relevance and quality criteria. These selected papers form the foundation for the subsequent analysis and review.

\section{For Finance: The Impact of Disclosing Carbon Emissions on Organizations}
\label{chp:impacts}

As global climate change worsens, corporate carbon emission management and transparent information disclosure have become critical topics of concern. Carbon emissions are not only vital indicators of a company's environmental footprint but also an essential measure of whether a company fulfills its sustainable development commitments. The disclosure of carbon emission information by organizations not only affects their market value but also has profound implications for the effectiveness of global carbon credit trading mechanisms. Given this context, this paper conducts an in-depth analysis of carbon emission disclosure practices in developed and developing countries, with a particular focus on the relationship between carbon emission disclosure and corporate market performance.

Studying carbon emission disclosure practices in these countries is of significant importance. First, differences in policies and market environments regarding carbon emission management and disclosure across countries can reveal the factors that drive or hinder organizations from disclosing carbon emission information. Second, an in-depth exploration of the impact of carbon emission disclosure helps understand how organizations can enhance market trust through transparent environmental responsibility practices and provide valuable insights for policymakers to further improve carbon credit trading mechanisms.

Recent studies in the United States suggest that firms with voluntary disclosure of carbon emissions experience higher valuations than their counterparts without voluntary disclosure. For example, a study conducted in the United States concluded that the market imposes stricter penalties on those who choose not to disclose relevant information, and that high carbon emissions have a negative impact on a company's market value \cite{carbon67}. This market dynamic is heavily reinforced by concrete regulatory developments in the U.S., where governmental bodies enforce strict penalties for non-compliance. For instance, at the operational level, regional emissions trading systems like California's Cap-and-Trade Program impose a stringent ``four-for-one'' compliance penalty, requiring organizations to surrender four allowances for every single allowance they are short \cite{narassimhan2018carbon}. Furthermore, on the disclosure and financial front, under the regulatory purview of the U.S. Securities and Exchange Commission (SEC), misleading or incomplete sustainability reporting can trigger severe administrative penalties and expose companies to costly shareholder litigation \cite{christensen2021mandatory}. These concrete policy mechanisms underscore that the threat of regulatory sanctions heavily bolsters the market's penalization of non-disclosure.

Similar results have been replicated in the United Kingdom. A study conducted in the United Kingdom, another developed country, shows that carbon emissions significantly negatively reduce corporate value, with markets typically reacting unfavorably to organizations with high carbon emissions. Even when organizations disclose more carbon emission information, the market response remains negative. This finding aligns with the results observed in the U.S. market \cite{carbon68}. Furthermore, the study has revealed a significant positive correlation between capital expenditure and carbon emission disclosure. Specifically, the more capital expenditure a company has, the more carbon emission information it discloses in its annual reports. This relationship is likely due to the fact that capital expenditure is typically associated with increased production activities, which in turn lead to higher carbon emissions \cite{carbon68}.

Additionally, the study highlights that robust internal corporate governance, such as board size, board independence, and gender diversity enhances the level of carbon emission disclosure \cite{carbon68}. Other studies focusing on UK organizations have explored the relationship between carbon emission, disclosure, and financial performance. The findings indicate a negative correlation between carbon emission levels and corporate financial performance. Conversely, a significant positive correlation exists between carbon information disclosure and financial performance, indicating that organizations that voluntarily disclose carbon information experience significant improvements in financial performance \cite{carbon77, carbon81}.

In the Australian market, disclosure policy plays a key role in influencing voluntary disclosure. The introduction of the National Greenhouse and Energy Reporting (NGER) Act in 2007 boosted voluntary disclosures of carbon emission, especially among larger, more visible firms. The research found that company size and carbon emission levels are key determinants of the extent of carbon emission disclosure. Larger organizations tend to disclose more carbon emission information, because they are more visible to the public and the regulators, thus having greater motivation to meet societal expectations through transparent disclosure practices. Furthermore, organizations with higher carbon emissions are more inclined to provide detailed disclosures, particularly in high-emission industries such as energy, transportation, materials, and utilities, where organizations face greater social and regulatory pressures, leading to more proactive and detailed disclosures \cite{carbon69}.

The study also suggests that good corporate governance is crucial in promoting carbon emission disclosure. Organizations with better governance structures (such as strong board independence and high governance scores) are more likely to disclose comprehensive carbon emission information, reflecting the positive role of good governance in enhancing corporate environmental transparency. Moreover, significant differences in disclosure practices exist across industries. Organizations in high-emission industries, due to greater social and political pressures, generally disclose more detailed carbon emission information, while in low-emission industries, better disclosure practices are often observed in larger organizations or those with more robust governance structures \cite{carbon69}.

Research in Japan further supports the aforementioned conclusions. Data analysis shows that reducing carbon emissions and disclosing carbon management activities have a positive impact on corporate market value. Specifically, the study reveals a significant negative correlation between corporate carbon emission levels and stock market value, while carbon management disclosure is positively correlated with stock market value. Additionally, the study finds that the positive correlation between carbon management disclosure and stock market value becomes more pronounced when carbon emission levels are higher. This suggests that reducing carbon emissions and actively disclosing related management activities are critical to a company’s sustainable development \cite{carbon73}.

In contrast to developed countries, studies conducted in developing countries have different results. For instance, a study in Indonesia found no direct effect of carbon emissions disclosure on firm value. Investors in these markets tend to prioritize financial performance over environmental performance. However, the study also found that disclosing carbon emission information can improve corporate financial performance, as organizations that disclose such information can attract environmentally conscious consumers, thereby increasing revenue and profitability, which is consistent with observations in developed countries \cite{carbon70}.

Further analysis suggests that effective government policy guidance is crucial for addressing climate change and enabling capital markets to hold organizations accountable for their carbon emissions. The implementation of mandatory disclosure policies, requiring organizations to publicly disclose their carbon emissions, is a key step in this process. Under such regulatory environments, the negative impact of carbon emissions on corporate market value becomes more significant, while voluntary disclosure of carbon emission information can partially mitigate this impact. It is through such mechanisms that the market can effectively penalize organizations with high carbon emissions and exert further pressure on those that choose not to disclose relevant information, which is why developed country governments generally adopt mandatory disclosure policies \cite{carbon70}.

For example, a study in Malaysia found that corporate carbon emission levels and disclosure do not significantly impact market value, further indicating that in the absence of mandatory policies, organizations lack motivation to strengthen carbon management \cite{carbon71}. Similar findings were observed in South Korea, where the impact of carbon emissions and disclosure on corporate market value is limited, as the capital market in Korea places greater emphasis on corporate profitability rather than environmental impact. Therefore, effective government policy guidance is particularly important, requiring stricter regulations to direct market attention towards corporate carbon emission behaviors \cite{carbon86}. In contrast, a study in the UK demonstrated that the government's mandatory disclosure requirements led to greater negative impacts on the market value of high-carbon-emission organizations, further confirming the critical role of policy in driving corporate carbon management and information disclosure \cite{carbon85}.

In addition to exploring the impact of carbon emissions on corporate market value, our research also found that carbon emission information disclosure and actual emission levels significantly influence corporate loan interest rates. The study indicates that organizations that voluntarily disclose carbon emission information typically enjoy lower loan spreads, particularly in organizations with lower information transparency (such as those with incomplete financial information or insufficient credit ratings). For these organizations, voluntary disclosure of carbon emission information can significantly reduce loan spreads, thereby lowering borrowing costs. Furthermore, the study reveals a positive correlation between actual corporate carbon emission levels, particularly direct carbon emissions (i.e., Scope 1 emissions), and loan spreads. This means that higher carbon emission levels lead to higher loan spreads. Financial markets, especially banks, consider environmental performance, particularly carbon emission levels, when assessing corporate credit risk. High carbon emissions are viewed as potential environmental risks, leading to an increase in loan spreads \cite{carbon72}.

This finding further underscores the importance of carbon emission information disclosure and carbon management. Transparent disclosure not only reduces negative impacts on the capital market but also brings tangible economic benefits to organizations by lowering loan costs.

Through a comprehensive analysis of carbon emission information disclosure research across multiple countries, we have drawn the following clear conclusions:

\begin{enumerate}
    \item \textbf{Impact of Carbon Emissions on Corporate Market Value:} In studies conducted in developed countries such as the United States, the United Kingdom, Australia, and Japan, a significant negative correlation was found between corporate carbon emission levels and market value. In other words, organizations with high carbon emissions tend to experience a negative impact on their market value. However, through voluntary disclosure of carbon emission information, organizations can partially mitigate this negative impact. This suggests that transparent carbon emission disclosure plays an important buffering role in capital markets.

    \item \textbf{Impact of Carbon Emission Disclosure on Corporate Financial Performance:} Research, particularly in the UK and Australia, shows a significant positive correlation between carbon emission disclosure and corporate financial performance. Organizations that voluntarily disclose carbon information tend to experience improved financial performance, possibly due to market recognition and support for their environmental responsibility.

    \item \textbf{Impact of Carbon Emission Disclosure on Bank Loan Interest Rates:} The study indicates that organizations that voluntarily disclose carbon emission information typically enjoy lower loan spreads, especially in organizations with lower information transparency. Conversely, higher carbon emission levels lead to increased loan spreads. This conclusion is applicable to both developed countries and some developing countries, indicating that financial institutions are increasingly considering corporate carbon emission behaviors when assessing risks.

    \item \textbf{The Key Role of Government Policy in Carbon Emission Disclosure:} Studies in developing countries (such as Indonesia and Malaysia) have shown that due to the lack of mandatory disclosure policies, carbon emission information disclosure has little impact on corporate market value. This highlights the crucial role of government policy guidance. In contrast, in countries like the UK with mandatory disclosure policies, the market value of high-carbon-emission organizations is more negatively impacted, further proving the critical role of policy in driving corporate carbon management.
\end{enumerate}

\begin{figure}[ht]
    \centering
    \begin{minipage}{0.48\textwidth}
        \centering
        \includegraphics[width=\linewidth]{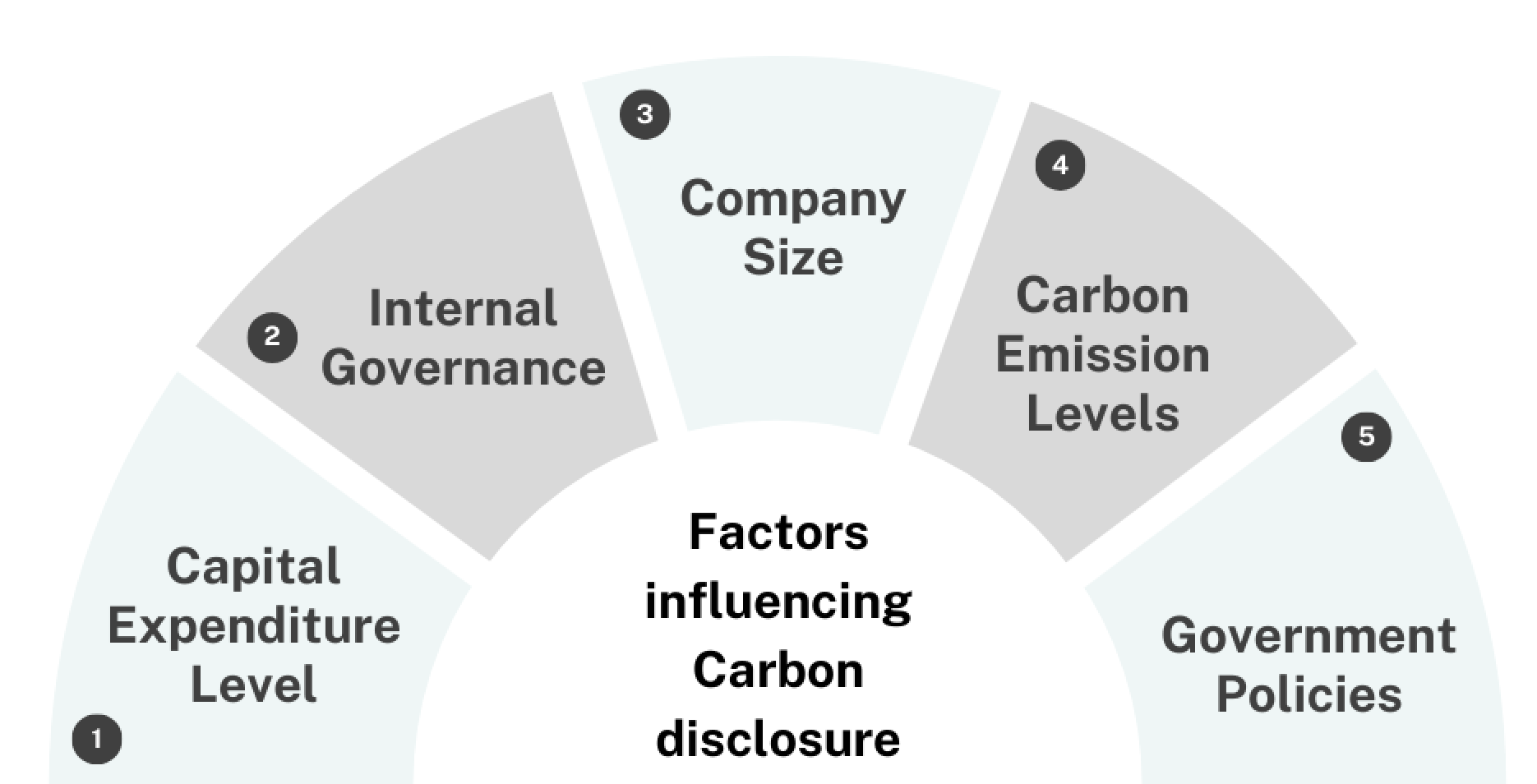}
        \caption{Factors influencing carbon emission disclosure}
        \label{fig:disclosure_factors}
    \end{minipage}
    \hfill
    \begin{minipage}{0.48\textwidth}
        \centering
        \includegraphics[width=\linewidth]{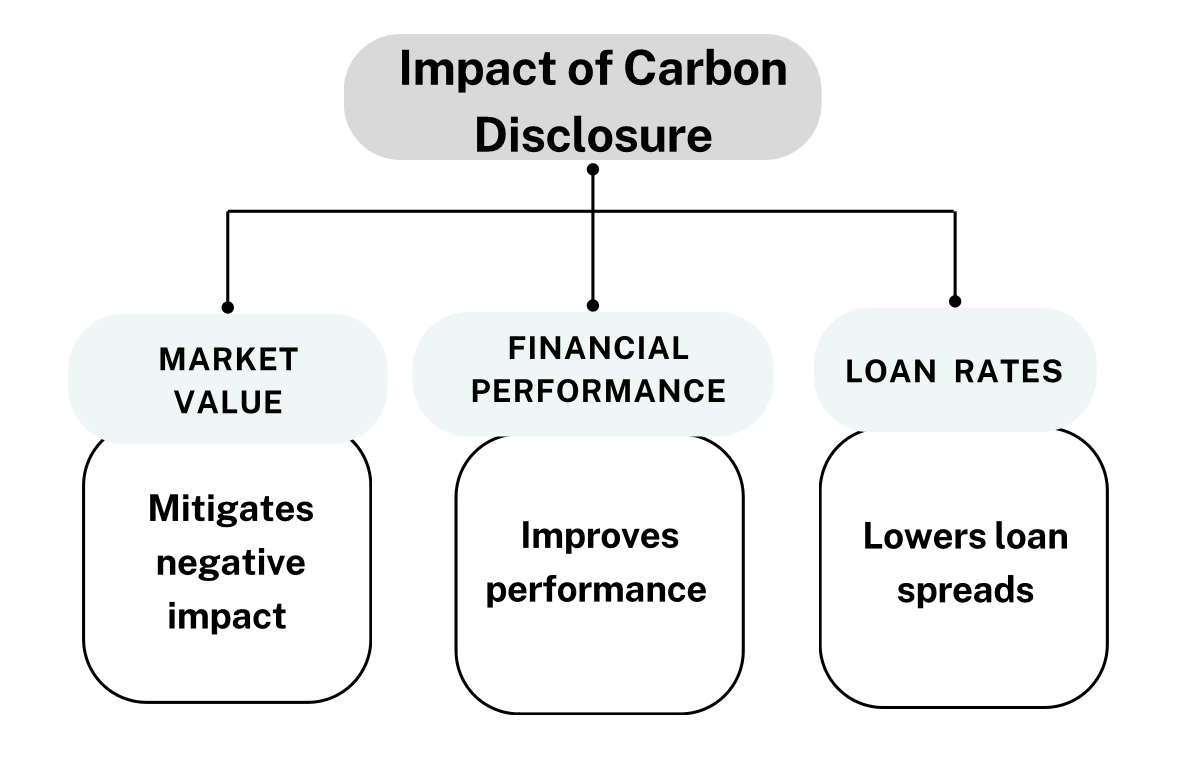}
        \caption{Impact of carbon emissions disclosure}
        \label{fig:Impact_of_disclosure}
    \end{minipage}
\end{figure}

Fig.~\ref{fig:disclosure_factors} and Fig.~\ref{fig:Impact_of_disclosure} summarize the factors influencing corporate carbon emission disclosure and the positive impacts of corporate carbon emission disclosure respectively.

In conclusion, organizations that engage in transparent carbon emission information disclosure and proactive carbon management can not only maintain a competitive edge in the market, gaining trust from capital markets and financial institutions, but also enhance financial performance, comply with government regulations, and reduce environmental risks, thereby promoting sustainable development. Therefore, organizations should integrate carbon management into their core strategies, actively disclose emissions, and pursue emission reductions to effectively respond to market and policy challenges, creating greater value for both themselves and society.

\section{For Engineering: Factors Influencing Carbon Credit Prices}
\label{chp:factors}

While transparent disclosure and proactive management are foundational for building market trust, relying solely on these internal practices is insufficient for a comprehensive carbon strategy. To further optimize their approaches, organizations must proactively engage with the carbon market, specifically by purchasing carbon credits when prices are favorable to reduce operational costs and maximize economic benefits. Research has found that for high-emission industries, carbon management costs will constitute a significant portion of expenses. Moreover, if carbon management costs are too high, the company’s competitive position will be weakened \cite{carbon81}. Therefore, understanding and analyzing the various factors that influence carbon credit prices is crucial. Through qualitative analysis of these factors, organizations can preliminarily predict carbon price trends, allowing them to more accurately determine the optimal timing for purchasing credits. Additionally, these factors can serve as input variables for carbon price prediction algorithms, enhancing the accuracy and applicability of predictions.

Next, this study undertakes a systematic review of the primary factors influencing carbon credit prices, as identified in existing literature, with the aim of providing valuable insights to support corporate decision-making in the carbon market. Table \ref{tab:factors-carbon-prices} provides a summary of the key factors that influence carbon credit prices and shows the experimental methods used to derive these conclusions. Table \ref{tab:carbon-market-products} offers a detailed breakdown of these influence factors across different carbon market products. The subsequent discussion synthesizes these findings and explores their implications.

\begin{footnotesize}
\setlength{\tabcolsep}{4pt}
\renewcommand{\arraystretch}{0.95}

\begin{longtable}{@{\extracolsep{\fill}}
  >{\raggedright\arraybackslash}p{0.14\textwidth}
  >{\raggedright\arraybackslash}p{0.36\textwidth}
  >{\raggedright\arraybackslash}p{0.16\textwidth}
  >{\raggedright\arraybackslash}p{0.27\textwidth}
@{}}
\caption{Summary of factors influencing carbon credit prices}
\label{tab:factors-carbon-prices}\\
\toprule
\textbf{Factor} & \textbf{Findings} & \textbf{Context / Mechanism} & \textbf{Methodology and Ref.} \\
\midrule
\endfirsthead

\toprule
\textbf{Factor} & \textbf{Findings} & \textbf{Context / Mechanism} & \textbf{Methodology and Ref.} \\
\midrule
\endhead

\midrule
\multicolumn{4}{r}{\textit{(Continued on next page)}}\\
\midrule
\endfoot

\bottomrule
\endlastfoot

\multirow[t]{2}{*}{Coal} &
Rising coal prices lead industries to seek alternative energy sources with fewer carbon emissions, like oil or renewable energy, reducing the demand for carbon credits and thus reducing carbon prices. &
\textbf{Substitution effect} (Fuel-switching to lower-carbon alternatives) &
Econometric Model \cite{n1} \newline
Static and Dynamic Panel Models \cite{n7} \newline
LASSO and SVAR Models \cite{n8} \newline
Multiscale Analysis Model \cite{n15} \newline
Vector Error Correction Model (VEC) and GARCH \cite{n17} \newline
OLS, IV, VAR model \cite{n18} \newline
Econometric Model \cite{n20} \newline
SVAR model \cite{n21} \\
\cmidrule(lr){2-4}
&
Rising coal prices are typically associated with increased demand for traditional energy sources, leading to higher carbon emissions and increased carbon prices. &
\textbf{Economic activity effect} (Production-driven emissions outpace fuel-switching) &
BP Neural Network Model \cite{n3} \newline
SEM and BN \cite{n5} \newline
Feasible Quasi Generalized Least Squares (FQGLS) \cite{n19} \\
\midrule

\multirow[t]{2}{*}{Crude Oil} &
Oil price increases are closely related to overall economic activity and have a strong positive impact on carbon prices. &
\textbf{Macroeconomic expansion} (Increased aggregate energy demand) &
Econometric Model \cite{n1} \newline
BP Neural Network Model \cite{n3} \newline
SV-TVP-VAR Model \cite{n4} \newline
SEM and BN \cite{n5} \newline
MEEMD and Fast Fourier Transform \cite{n6} \newline
TVP-VAR Model \cite{n10} \newline
Feasible Quasi Generalized Least Squares \cite{n19} \newline
SVAR model \cite{n21} \\
\cmidrule(lr){2-4}
&
Increases in crude oil prices raise production costs, leading organizations to reduce supply, which lowers carbon emissions and reduces the demand and price of carbon emission futures. &
\textbf{Cost-push constraint} (Production scale reduction) &
Vector Error Correction Model \cite{n2} \newline
SV-TVP-VAR Model \cite{n4} \\
\midrule

Macroeconomics &
Growth in macroeconomic activity typically drives increases in industrial production and raw material consumption, leading to higher demand for carbon emissions and pushing carbon prices up. &
\textbf{Industrial demand expansion} (Activity effect) &
Econometric Model \cite{n1} \newline
Vector Error Correction Model \cite{n2} \newline
BP Neural Network Model \cite{n3} \newline
MEMD, VAR, VEC Models \cite{n14} \newline
Ensemble Empirical Mode Decomposition (EEMD) \cite{n16} \newline
Vector Error Correction Model and GARCH \cite{n17} \newline
SVAR model \cite{n21} \\
\midrule

Similar Products &
An increase in similar products' prices often leads to a rise in carbon credit prices, indicating a substitution relationship between the two. &
\textbf{Substitution effect} (Cross-market hedging) &
Econometric Model \cite{n1} \newline
SV-TVP-VAR Model \cite{n4} \newline
ARMA-GARCH \cite{n12} \\
\midrule

Stock Market Indices &
Stock market indices have a significant positive impact on the carbon market. &
\textbf{Market sentiment \& Economic expectations} &
SEM and BN \cite{n5} \newline
Static and Dynamic Panel Models \cite{n7} \newline
LASSO and SVAR Models \cite{n8} \newline
ARIMA-GARCH Model, Random Forest Algorithm \cite{n13} \newline
Multiscale Analysis Model \cite{n15} \\
\midrule

Euro Index &
The Euro Index had a negative impact on carbon prices before COVID-19, mainly due to uncertainties like Brexit. &
\textbf{Geopolitical \& Economic uncertainty} &
MEEMD and Fast Fourier Transform \cite{n6} \\
\midrule

Utility Index &
The utility index positively impacts carbon prices, reflecting increased power plant activity and associated carbon emissions. &
\textbf{Sector-specific emission demand} (Power generation) &
MEEMD and Fast Fourier Transform \cite{n6} \\
\midrule

Interbank lending rates &
Interbank lending rates negatively correlate with carbon prices; rising rates weaken market liquidity and reduce demand for carbon allowances, leading to lower carbon prices. &
\textbf{Liquidity channel} (Tightened capital availability) &
Bai-Perron Structural Break Test and Newey-West Regression Model \cite{n9} \\
\midrule

Exchange rate &
Exchange rates negatively impact carbon prices, particularly in the short and medium term. &
\textbf{Trade competitiveness \& Capital flow} &
DCC-GARCH, TVP-VAR and GA-BP \cite{n11} \newline
MEMD, VAR, VEC Models \cite{n14} \\
\midrule

COVID-19 Pandemic &
The COVID-19 pandemic significantly reduced carbon prices. &
\textbf{Macroeconomic \& Supply chain shock} &
Bai-Perron Structural Break Test and Newey-West Regression Model \cite{n9} \newline
EEMD \cite{n16} \\
\midrule

Clean Energy &
Rising clean energy prices will drive up carbon prices, while the development of clean energy may sometimes suppress carbon prices. &
\textbf{Substitution effect vs. Technology adoption} &
BP Neural Network Model \cite{n3} \newline
SEM and BN \cite{n5} \newline
MEEMD and Fast Fourier Transform \cite{n6} \newline
Static and Dynamic Panel Models \cite{n7} \newline
TVP-VAR Model \cite{n10} \newline
OLS, IV, VAR model \cite{n18} \newline
Feasible Quasi Generalized Least Squares \cite{n19} \newline
Econometric Model \cite{n20} \newline
SVAR model \cite{n21} \\
\midrule

Electricity price &
Electricity prices positively influence the price of carbon credits. &
\textbf{Energy market linkage} (Cost pass-through) &
TVP-VAR Model \cite{n10} \newline
Multiscale Analysis Model \cite{n15} \newline
OLS, IV, VAR model \cite{n18} \newline
Econometric Model \cite{n20} \\
\midrule

End product prices &
The rise in prices of end products such as minerals, steel, and paper can drive up carbon credit prices, as the increase in end product prices leads to higher production, thereby increasing the demand for carbon emissions. &
\textbf{Derived demand} (Production expansion) &
OLS, IV, VAR model \cite{n18} \\
\midrule

Air quality &
Environmental factors like air quality significantly impact carbon prices; higher air quality index (AQI) levels are associated with rising carbon prices. &
\textbf{Environmental indicator proxy} (High emission periods) &
ARMA-GARCH \cite{n12} \newline
VAR-VEC model and GARCH \cite{n22} \newline
Vector Error Correction Model and GARCH \cite{n17} \\
\midrule

Temperature &
Temperature has a significant impact on carbon allowance trading volume. Extreme weather conditions, such as very cold or hot temperatures, increase energy demand, thereby affecting the carbon allowance market and driving up carbon prices. &
\textbf{Energy demand spike} (Heating/Cooling needs) &
Econometric Model \cite{n20} \newline
Autoregressive Distributed Lag Model (ARDL) \cite{n23} \newline
Structural Break Test \cite{n24} \\
\midrule

Supply factor &
The supply of carbon credits also affects carbon prices; when the supply is large, carbon prices decrease. &
\textbf{Market equilibrium} (Supply surplus) &
Econometric Model \cite{n20} \\
\midrule

Degree of Carbon Regulation &
Short-term negative impact of carbon regulation policies on carbon prices is greater than long-term effects, but carbon prices become more sensitive to policy information as market mechanisms improve. &
\textbf{Policy shock \& Compliance cost sensitivity} &
SV-TVP-VAR Model \cite{n4} \\
\midrule

\multirow[t]{2}{*}{Interest Rates} &
Interest rates have a negative short-term impact on carbon prices. &
\textbf{Short-term: Liquidity channels} (High opportunity cost) &
SV-TVP-VAR Model \cite{n4} \\
\cmidrule(lr){2-4}
&
Interest rates have a positive impact on carbon prices, especially in the medium to long term, as higher interest rates increase the cost of capital for green transition projects, sustaining reliance on high-emission production. &
\textbf{Medium/Long-term: Investment barriers} (Cost of capital) &
DCC-GARCH, TVP-VAR and GA-BP \cite{n11} \\
\midrule

Gold Prices &
Gold prices positively affect carbon prices, indicating a relationship between increased gold mining activities and higher carbon emissions. &
\textbf{Mining activity emissions} (Energy intensive) &
MEEMD and Fast Fourier Transform \cite{n6} \\
\midrule

Population growth rate &
There is a positive correlation between population growth rate and carbon prices. A higher population growth rate will increase future carbon prices, as population growth leads to higher carbon emissions. &
\textbf{Demographic scale effect} (Absolute demand growth) &
EZ Climate Model and Uncertainty Analysis \cite{n25} \\
\midrule

Growth rate of per capita output &
The growth rate of per capita output has a greater impact on carbon prices. A higher per capita output growth rate will significantly increase future carbon prices. &
\textbf{Economic scale effect} (Per capita consumption) &
EZ Climate Model and Uncertainty Analysis \cite{n25} \\

\end{longtable}
\end{footnotesize}

A systematic review of existing literature reveals that carbon credit prices are influenced by a variety of factors, such as coal and crude oil. These factors include not only fluctuations in traditional energy markets but also macroeconomic indicators, regulatory policies, market sentiment, and external shocks. The following provides a detailed summary of these factors and their mechanisms of impact.

Energy prices play a crucial role in determining carbon credit prices. Increases in coal and crude oil prices typically lead to higher carbon emissions, which in turn drive up carbon credit prices. However, as coal prices rise, industrial sectors may shift towards alternative energy sources with lower carbon emissions, such as natural gas or renewable energy, which could reduce the demand for carbon credits and subsequently lower prices. The impact of natural gas prices on carbon credits can be twofold too. On one hand, higher natural gas prices may elevate carbon credit prices. On the other hand, the development of renewable energy may temper this upward trend \cite{n1,n3,n4,n5,n6,n7,n8,n10,n15}.

The dual impact of coal prices on carbon credit valuations can be attributed to the interplay between the \textit{substitution effect} and the \textit{economic activity effect}. On one hand, when coal prices rise relative to cleaner alternatives like natural gas, a substitution effect occurs: industrial producers shift to lower-carbon fuels, thereby depressing the demand for carbon allowances and leading to a price decline. On the other hand, a surge in coal prices often mirrors periods of robust global industrial expansion and heightened energy consumption. In such scenarios, the activity effect dominates, where the overall increase in production-driven emissions outpaces fuel-switching efforts, ultimately exerting upward pressure on carbon prices.

Macroeconomic indicators and stock market performance also significantly influence carbon credit prices. Economic growth often accompanies increased industrial production and energy consumption, thereby raising the demand for carbon emissions and driving up carbon credit prices. Moreover, stock market indices, such as the Euro Stoxx 50 and S\&P 500, play a critical role in the carbon market, particularly during periods of economic prosperity. However, in certain markets, such as China, the relationship between stock market performance and carbon prices may be negative, indicating that market maturity and investor expectations influence price fluctuations \cite{n1,n2,n3,n5,n7,n8,n13,n14,n15,n16}.

Regulatory policies have a significant impact on the carbon market. In the short term, stringent carbon regulation policies may increase compliance costs for organizations, thereby depressing carbon prices. However, as market mechanisms improve, the sensitivity of carbon prices to policy information may increase, potentially leading to positive impacts in the long run. Additionally, the price linkage between international carbon markets is another key consideration, particularly the positive impact of European carbon markets on China's carbon market, which has become more evident since 2020 \cite{n4,n7,n9,n10}.

Financial factors such as exchange rates and interest rates also have complex effects on carbon credit prices. Exchange rate fluctuations generally have a negative impact on carbon prices in the short and medium term. For example, when the Renminbi depreciates, carbon prices may rise in the short term but decline in the long term. Rising interest rates typically suppress market liquidity, reducing demand for carbon credits and leading to lower prices. However, in the medium to long term, rising interest rates may increase the cost of capital for green projects, delaying emission reductions and thereby sustaining the demand for carbon credits, which pushes up carbon prices. \cite{n9,n11,n14}. These conclusions have also been validated in several studies. For example, \cite{n9} justifies the negative relationship between interbank lending rates and carbon prices using the Bai-Perron Structural Break Test and Newey-West Regression Model. Similarly, the complex influence of interest rates reflects the transition from short-term \textit{liquidity channels} to long-term \textit{investment barriers}. In the short term, higher interest rates typically tighten market liquidity and increase the opportunity cost of holding carbon assets, which incentivizes sell-offs and drives prices down. Conversely, in the medium to long term, elevated interest rates significantly increase the cost of capital for renewable energy and carbon-mitigation projects, which are characteristically capital-intensive. By creating a financial barrier to the green transition, high interest rates can delay the implementation of emission-reduction technologies, leading to a persistent reliance on traditional high-emission processes and a sustained demand for carbon credits in the long run.

Market sentiment and significant events, such as the COVID-19 pandemic, have profound effects on carbon credit price volatility. During the pandemic, global economic uncertainty and supply chain disruptions led to a sharp decline in carbon prices. However, the implementation of the European Union's 750 billion euro green recovery plan helped stabilize the carbon market, leading to a rebound in carbon prices. Additionally, fluctuations in market sentiment may exhibit nonlinear relationships, adding complexity to price predictions. In such cases, nonlinear models like random forests have demonstrated superior accuracy in predicting price changes \cite{n6,n9,n11,n12,n1}.

Air quality is a significant environmental factor that influences carbon prices. Higher Air Quality Index levels are typically associated with rising carbon prices, as poorer air quality often indicates increased carbon emissions, thereby driving up carbon prices. Several studies have confirmed this relationship \cite{n12,n22,n17}. For instance, \cite{n12} validated this positive correlation using the ARMA-GARCH model.

Temperature has a notable impact on carbon allowance trading volumes. Extreme weather conditions, such as very cold or hot temperatures, increase energy demand, which in turn affects the carbon allowance market and drives up carbon prices. Multiple studies have validated this effect using various econometric models \cite{n20,n23,n24}.

The supply of carbon credits also affects carbon prices. When the supply of carbon credits is large, prices tend to decrease due to market oversupply. \cite{n20} demonstrated the inverse relationship between supply levels and carbon prices using an econometric model.

Gold prices have a positive effect on carbon prices, as increased gold mining activities are typically associated with higher carbon emissions, which raises the demand for carbon credits. \cite{n6} used the MEEMD and Fast Fourier Transform models to demonstrate the positive relationship between gold prices and carbon prices.

There is a positive correlation between the population growth rate and carbon prices. A higher population growth rate is expected to increase future carbon prices, as population growth is often accompanied by higher carbon emissions. \cite{n25} used the EZ Climate Model and Uncertainty Analysis to demonstrate the impact of population growth on carbon prices.

The growth rate of per capita output significantly influences carbon prices. Higher per capita output growth rates will lead to substantial increases in future carbon prices, as this growth is typically accompanied by increased industrial activity and energy consumption. \cite{n25} validated the influence of per capita output growth on carbon prices using the EZ Climate Model and Uncertainty Analysis.

In addition to the factors mentioned above, other influences have also been shown to affect carbon prices. For example, \cite{n12} justifies the negative relationship between the prices of similar products using the ARMA-GARCH model. Electricity prices have also been confirmed by numerous studies to have a positive impact on carbon prices \cite{n10,n15,n18,n20}.

In conclusion, carbon credit price fluctuations result from the interplay of various factors. Energy prices, macroeconomic activity, regulatory policies, financial market dynamics, market sentiment, and external events collectively shape the dynamics of the carbon credit market. Understanding the interactions among these factors is crucial for predicting carbon price trends and optimizing corporate carbon management strategies.

\begin{footnotesize}
\setlength{\tabcolsep}{4pt}
\renewcommand{\arraystretch}{0.95}

\begin{longtable}{@{\extracolsep{\fill}}
  >{\raggedright\arraybackslash}p{0.22\textwidth}
  >{\raggedright\arraybackslash}p{0.48\textwidth}
  >{\raggedright\arraybackslash}p{0.30\textwidth}
@{}}
\caption{Key Factors Influencing Carbon Markets Across Regions}
\label{tab:carbon-market-products}\\
\toprule
\textbf{Market} & \textbf{Factor} & \textbf{Reference} \\
\midrule
\endfirsthead

\toprule
\textbf{Market} & \textbf{Factor} & \textbf{Reference} \\
\midrule
\endhead

\midrule
\multicolumn{3}{r}{\textit{(Continued on next page)}}\\
\midrule
\endfoot

\bottomrule
\endlastfoot

\multirow[t]{6}{*}{\makecell[tl]{European Union\\Allowance\\(EUA)}} &
Black Energy (Coal, Oil) and Clean Energy (Natural Gas, Solar) &
BP Neural Network Model \cite{n3} \\
\cmidrule(lr){2-3}
&
Gold Price, Euro Index and Utility Index &
MEMD and FFT \cite{n6} \\
\cmidrule(lr){2-3}
&
Stock Market Indices (SP 500, CAC 40, DAX) and Carbon Emission Market (EUA, CER) &
SEM and BN \cite{n5} \\
\cmidrule(lr){2-3}
&
Market Stability Reserve (MSR) &
N-A MEMD, VAR, VEC Models \cite{n14} \\
\cmidrule(lr){2-3}
&
Energy Market and Stock Market &
LASSO and MSVAR Models \cite{n8} \\
\cmidrule(lr){2-3}
&
COVID-19 Pandemic &
Bai-Perron Structural Break Test and Newey-West Regression Model \cite{n9} \\
\midrule

\multirow[t]{6}{*}{China's Carbon Market} &
Degree of Carbon Regulation, Crude Oil Price, EUA Futures Price and Interest Rates &
SV-TVP-VAR Model \cite{n4} \\
\cmidrule(lr){2-3}
&
Exchange Rate and Coal Price &
DCC-GARCH, TVP-VAR and GA-BP \cite{n11} \\
\cmidrule(lr){2-3}
&
Industrial Production, Stock Market Index and Renewable Energy Generation &
Static and Dynamic Panel Models \cite{n7} \\
\cmidrule(lr){2-3}
&
Industrial Structure, Policy Factors, Environmental Factors and External Market &
ARMA-GARCH \cite{n12} \\
\cmidrule(lr){2-3}
&
Energy Prices, Economic Indicators, Macroeconomic Variables and Market Sentiment &
ARIMA(1,1,1)-GARCH(1,1) Model, Random Forest Algorithm \cite{n13} \\
\cmidrule(lr){2-3}
&
Significant Short-term Events &
EEMD \cite{n16} \\

\end{longtable}
\end{footnotesize}

Table \ref{tab:carbon-market-products} summarizes the key factors influencing the European Union Allowance market and China’s carbon market, along with the relevant research models and methodologies. This table clearly illustrates the differences in influencing factors across these markets.

In the EUA market, carbon prices are affected by energy prices, financial market indicators, policy factors, and external events. In terms of the energy market, price fluctuations of black energy (such as coal and crude oil) and clean energy (such as natural gas and solar) are major drivers of carbon prices. As traditional energy prices rise, organizations tend to shift toward low-carbon energy, thereby affecting the supply and demand of carbon credits \cite{n3}. Regarding financial market indicators, factors such as gold prices, the Euro Index, and the utility index also influence carbon prices \cite{n5,n6,n8}. These indicators reflect macroeconomic and financial market changes, and their connection to the energy market has been verified through models such as MEMD and FFT \cite{n6}. As for stock market and carbon market indices, stock market indices (such as SP 500, CAC 40, DAX) have a significant positive impact on the carbon market, as they reflect economic growth expectations, which in turn affect energy demand and carbon emissions \cite{n5,n8}. Additionally, the Market Stability Reserve (MSR) is a policy mechanism unique to the EUA market, designed to stabilize the market and prevent excessive carbon price fluctuations, with its impact analyzed through VAR and VEC models \cite{n14}. The COVID-19 pandemic, as an external event, caused short-term shocks to the carbon market, most notably the sharp decline in carbon prices in 2020, followed by a recovery driven by stimulus plans \cite{n9}.

In contrast, China's carbon market is influenced by more policy and structural factors. Carbon regulation intensity is a critical factor in China’s carbon market, as the implementation of carbon regulation policies has a direct short-term impact on carbon prices, particularly by increasing compliance costs for organizations and suppressing market demand \cite{n4}. Crude oil and coal prices also play a crucial role in China’s market, with coal price fluctuations being especially prominent, reflecting China’s reliance on coal in its energy structure \cite{n11}. In addition, exchange rates and interest rates significantly impact China’s carbon market, particularly in the short and medium term, as fluctuations in exchange rates and interest rates affect market liquidity and energy prices \cite{n11,n4}. Lastly, the growth of industrial production and renewable energy also has a significant influence on China’s carbon prices. These factors reflect the level of industrialization in China and its ongoing energy transition, with increases in industrial production and renewable energy generation being closely linked to carbon market prices \cite{n7,n16}.

Table \ref{tab:factors-carbon-prices} and Table \ref{tab:carbon-market-products} together provide valuable reference tools for future researchers, helping them gain a comprehensive understanding of the key factors influencing carbon credit prices and the analytical methods used to examine them. Table \ref{tab:factors-carbon-prices} summarizes the critical factors affecting carbon credit prices and presents the experimental methods used to derive these conclusions, helping researchers identify broadly applicable variables and validation models for the carbon market. Table \ref{tab:carbon-market-products} further refines this by demonstrating how these influencing factors manifest across different carbon market products, enabling researchers to select appropriate analytical directions and models based on the characteristics of their target markets.

\section{Algorithms for Predicting Carbon Credit Prices}
\label{chp:algorithmsCC}

After thoroughly reviewing the factors influencing carbon prices, it becomes clear that these factors are not only critical for understanding carbon market fluctuations but also serve as essential input variables for carbon price prediction algorithms. Effectively incorporating these factors into predictive models can significantly enhance the accuracy and practical value of carbon price forecasts, making the review of these algorithms both meaningful and necessary. By accurately predicting carbon prices, organizations can strategically plan their carbon credit purchases when prices are lower, thereby better managing their costs. This not only helps organizations control overall expenditures and strengthen their competitiveness in the market but also encourages them to actively participate in carbon management and transparently disclose their carbon emissions. In doing so, organizations can gain greater trust and advantage in markets that increasingly prioritize ESG principles. The following section reviews recent carbon price prediction algorithms and their methodological development.

Before delving into specific empirical models, it is crucial to establish the theoretical motivation driving the methodological progression in carbon price forecasting. Unlike traditional financial assets, carbon credit markets exhibit profound non-linearity, high volatility, and non-stationarity. These characteristics stem from their unique susceptibility to macroeconomic shifts, energy price fluctuations, and, most importantly, sudden policy interventions and quota adjustments. Traditional empirical models or single machine learning algorithms, such as basic SVR or ARIMA, often fail to capture these multi-scale, complex dynamics, as they struggle to differentiate between long-term policy trends and short-term market noise.

Consequently, the literature has increasingly gravitated toward a ``decomposition-and-optimization'' paradigm. The theoretical rationale for this shift is rooted in signal processing: by employing decomposition techniques, such as EMD, CEEMDAN, and VMD, complex pricing signals can be isolated into more predictable and relatively stationary intrinsic mode functions (IMFs). Subsequently, optimization algorithms, such as PSO, GA, and GWO, are introduced to dynamically tune the hyperparameters of neural networks or other forecasting models for each specific frequency component, thereby improving predictive accuracy. This evolution from static modeling to dynamic, multi-stage hybrid architectures represents a necessary adaptation to handle the inherent complexity of the carbon market.

To make this methodological progression clearer, the following review is organized according to the dominant methodological contribution of each study rather than as a purely chronological list. The reviewed forecasting models are grouped into four categories: foundational decomposition and shallow/kernel forecasting models, optimized decomposition-ensemble and deep point forecasting models, uncertainty-aware and leakage-aware forecasting models, and deployment-oriented advanced architectures with interpretability, robustness, and external information. This organization preserves the technical details of individual studies while clarifying how carbon price forecasting methods have evolved from decomposition-based point prediction toward more risk-aware and deployment-oriented FinTech applications.

\medskip
\noindent\textbf{Foundational decomposition and shallow/kernel forecasting models.}

This group of studies establishes the basic decomposition-based forecasting paradigm in carbon price prediction. These models mainly use signal decomposition or filtering techniques to reduce non-stationarity and noise before applying shallow, kernel-based, or relatively lightweight learning models to forecast carbon price movements.

Sun and Wang used a combined method of empirical mode decomposition \cite{38emd} and least squares support vector machine to predict carbon prices. They proposed a hybrid model based on factor analysis, empirical mode decomposition, improved particle swarm optimization \cite{38swamoptim}, and least squares support vector machine \cite{LSVR38} to consider historical carbon prices and external factors influencing carbon prices. The study investigated three typical carbon markets in China, and the results showed that this hybrid model was more accurate in predicting carbon prices than other models \cite{carbon38}.

Hao and Tian proposed an innovative carbon price prediction framework. First, the framework applies ICEEMDAN \cite{ICEEMDAN40} to decompose the carbon price time series, effectively removing noise while retaining key signal features. Then, feature selection is conducted using the Partial Autocorrelation Function (PACF) and Maximum Relevance Minimum Redundancy (mRMR) methods. Next, the parameters of the KELM model are optimized using the Multi-Objective Chaotic Sine-Cosine Algorithm (MOCSCA). Finally, a multi-step prediction strategy is designed using the recursive method, which not only accurately predicts short-term price fluctuations but also provides reliable support for long-term trend forecasting \cite{carbon40}.

Liu and Shen proposed a carbon price prediction model based on Empirical Wavelet Transform (EWT) \cite{EWT43} and Gated Recurrent Unit, aiming to enhance prediction accuracy by integrating multiple techniques. First, the carbon price time series is decomposed using EWT, extracting components of different frequencies, including trend components, low-frequency components, and high-frequency components. Subsequently, Fuzzy C-Means (FCM) \cite{FCM43} clustering is employed to classify these components, further refining the signal characteristics. Then, the Partial Autocorrelation Function is used to analyze the lag relationships of each component, determining the optimal lag order as input for the GRU model. Finally, the GRU model performs single-step predictions for each component, and the predicted results are aggregated to obtain the overall carbon price prediction \cite{carbon43}.

Huang and He proposed a carbon price prediction model based on multi-scale joint prediction, integrating techniques such as Mean Optimization Empirical Mode Decomposition (MOEMD), Grey Relational Analysis (GRA), Factor Analysis (FA), Cooperative Kidney Algorithm (CKA) \cite{cka44}, and Extreme Learning Machine (ELM). First, MOEMD is used to decompose the carbon price data, extracting modal components at different frequencies. Then, GRA and FA are employed to select both structured and unstructured data highly correlated with carbon price fluctuations, including unstructured data like Baidu Search Index. Subsequently, CKA is utilized to optimize the parameters of the ELM model, enhancing prediction accuracy and stability. The experimental results demonstrate that the proposed MOEMD-CKA-ELM model significantly outperforms other comparative models in carbon price prediction \cite{carbon44}.

Hao et al. proposed a novel hybrid model based on feature selection and multi-objective optimization algorithms for carbon price prediction. The study used Weighted Regularized Extreme Learning Machine (WRELM) as the artificial intelligence algorithm to improve the accuracy and stability of carbon price prediction \cite{wrema52}. The model overcame the limitations of statistical models in handling carbon price prediction and optimized the model using the Multi-Objective Grasshopper Optimization Algorithm (MOGOA), achieving excellent prediction results \cite{mogoa53}. This research provides important references for the rational and stable development of the carbon market \cite{carbon48}.

Zhao et al. proposed a carbon price prediction model that combines the Hodrick-Prescott (HP) filter, an improved Grey Model (Variable Grey Model, VGM), and Extreme Learning Machine \cite{vgm57}. First, the HP filter is used to decompose the carbon price data and its influencing factors into long-term trends and short-term fluctuations, which are modeled separately. The VGM model is then employed to predict the long-term trends, while the ELM handles the short-term fluctuations. By combining the predictions from these two components, the model effectively enhances the accuracy of carbon price prediction \cite{carbon51}.

Taken together, this group demonstrates the early methodological foundation of carbon price forecasting: decomposition or filtering is first used to stabilize nonlinear and non-stationary price signals, and then relatively lightweight forecasting models are applied to the processed components. As later summarized in Table~\ref{tab:eval_matrix}, these studies mainly evaluate performance through conventional point-error metrics such as RMSE, MAE, and MAPE, with only limited use of formal statistical testing or dynamic validation designs. Their main contribution is therefore to validate the usefulness of decomposition-based preprocessing for carbon price prediction. However, because most of these models rely on static train-test splits, their evidence for real-time deployment and robustness under changing carbon market regimes remains limited.

\medskip
\noindent\textbf{Optimized decomposition-ensemble and deep point forecasting models.}

Building on the foundational decomposition-based models, the second group of studies further improves point forecasting accuracy by introducing secondary decomposition, ensemble learning, deep learning architectures, and heuristic optimization. Compared with the first group, these models usually contain longer modeling pipelines and more intensive parameter tuning, with the main objective of reducing deterministic prediction errors.

Nadirgil employed the Complete Ensemble Empirical Mode Decomposition with Adaptive Noise (CEEMDAN) \cite{CEEMDAN39} technique to decompose the carbon price time series into multiple Intrinsic Mode Functions (IMFs), capturing various frequency characteristics within the data. Subsequently, Permutation Entropy (PE) was utilized to classify the IMFs, grouping IMFs with similar complexity into Cooperative Intrinsic Mode Functions (Co-IMFs) to reduce model computational complexity and the risk of overfitting. For high-frequency Co-IMFs, Variational Mode Decomposition (VMD) was further applied for secondary decomposition to mitigate the impact of high-frequency noise on model prediction. Ultimately, the study integrated multiple machine learning models, including Long Short-Term Memory (LSTM), Gated Recurrent Unit (GRU), Multilayer Perceptron (MLP), and Backpropagation Neural Network (BPNN), and optimized model parameters using Genetic Algorithm (GA) to construct 48 different hybrid prediction models. The effectiveness of the dual decomposition method (CEEMDAN-VMD-BPNN-GA) in addressing prediction complexity was ultimately demonstrated \cite{carbon39}.

Li and Liu proposed a carbon price prediction model based on secondary decomposition and feature selection. First, the carbon price time series is subjected to multi-level decomposition using Discrete Wavelet Transform (DWT) \cite{DWT41} and ICEEMDAN, breaking it down into sub-sequences of different frequencies. Simultaneously, a three-stage feature selection method is employed to identify external influencing factors that are highly correlated with carbon prices. Subsequently, Support Vector Regression (SVR) \cite{SVR41} and Multilayer Perceptron are used to predict the high-frequency and low-frequency sub-sequences, respectively, in order to capture short-term fluctuations and long-term trends. Finally, a weighted ensemble of the predictions from each sub-sequence is performed to obtain the overall carbon price prediction \cite{carbon41}.

Yang et al. proposed a carbon price prediction model based on a multi-factor combination, integrating Modified Ensemble Empirical Mode Decomposition (MEEMD), Long Short-Term Memory networks \cite{lstm45}, and a rule-based machine reasoning system. First, MEEMD is employed to decompose the carbon price time series, extracting modal components of different frequencies, which provide clear signals for subsequent predictions \cite{meemd46}. Then, the LSTM model, with its strong capability for processing time series data, accurately predicts each modal component. To further enhance prediction accuracy, a rule-based machine reasoning system is introduced to automatically optimize the key parameters of the LSTM model, ensuring robust performance under various conditions. The experimental results demonstrate that the proposed MEEMD-LSTM model outperforms traditional models in carbon price prediction \cite{carbon45}.

Lu et al. proposed a carbon price and trading volume prediction framework based on multi-model comparison, incorporating the Complete Ensemble Empirical Mode Decomposition with Adaptive Noise technique to evaluate the performance of six different machine learning models. The machine learning models used include XGBoost \cite{xgboost47}, Random Forest \cite{random48}, Grey Wolf Optimizer-based Nonlinear Extension of the Arps Decline Model (GWO-KNEA) \cite{knea49}, Particle Swarm Optimization-based Support Vector Machine (PSO-SVM) \cite{psosvm50}, Simulated Annealing and Fruit Fly Optimization Algorithm-based Support Vector Machine (SA-FFOA-SVM), and Radial Basis Function Neural Network (RBFNN) \cite{rbfnn51}. Through experiments conducted on data from eight carbon trading markets in China, the authors compared the prediction accuracy and stability of each model. The study found that the CEEMDAN-GWO-KNEA and CEEMDAN-RBFNN models performed exceptionally well across multiple datasets \cite{carbon47}.

Yang et al. proposed a carbon price prediction model based on Modified Ensemble Empirical Mode Decomposition, Long Short-Term Memory networks, and an Improved Whale Optimization Algorithm (IWOA) \cite{iwoa56}. First, MEEMD is used to decompose the carbon price time series into several Intrinsic Mode Functions and a residual component, reducing noise and mode mixing issues \cite{meemd54}. Subsequently, the LSTM model is employed to predict the future values of each IMF and the residual component, capturing long-term dependencies in the time series \cite{lstm45}. To further enhance prediction performance, IWOA is used to optimize the initial weights and biases of the LSTM model \cite{carbon49}.

Zhang et al. proposed a hybrid model for multi-step carbon price prediction, integrating Extremely Randomized Trees (ET), Multivariate Variational Mode Decomposition (MVMD), and Long Short-Term Memory networks. The model employs ET to select the most influential external factors affecting carbon price fluctuations, simplifying the model's complexity and enhancing prediction accuracy. MVMD \cite{mvmd65} is then used to decompose the carbon price and selected external variables into multiple modes, allowing for the extraction of stable frequency components that are crucial for accurate predictions. Each mode is subsequently modeled using LSTM to capture long-term dependencies in the time series, with the final prediction obtained by aggregating the individual predictions \cite{np14}.

Overall, this group represents the transition from basic decomposition-based forecasting to more complex optimized hybrid point forecasting. These studies show that secondary decomposition, multi-model comparison, feature screening, and heuristic optimization can further improve deterministic forecasting accuracy, especially for highly nonlinear and noisy carbon price series. However, as later summarized in Table~\ref{tab:eval_matrix}, this improvement is usually accompanied by longer modeling pipelines, more hyperparameters, and greater dependence on static experimental settings. Therefore, although these models report stronger point prediction performance than many simpler baselines, their real-world deployment still requires careful attention to overfitting risk, tuning burden, and possible decomposition-related leakage.

\medskip
\noindent\textbf{Uncertainty-aware and leakage-aware forecasting models.}

Beyond deterministic point forecasting, the third group of studies extends carbon price prediction toward interval prediction, probabilistic forecasting, and more rigorous handling of time-series preprocessing risks. These models are particularly relevant to FinTech applications because carbon credit trading, hedging, budgeting, and asset valuation require not only predicted price levels but also uncertainty ranges and risk information.

Wang et al. proposed a comprehensive carbon price prediction framework. Initially, the carbon price time series undergoes multi-level decomposition and denoising using Complete Ensemble Empirical Mode Decomposition with Adaptive Noise and Wavelet Transform (WT) to extract key features and reduce noise interference. Subsequently, Principal Component Analysis (PCA), Random Forest (RF), and Gradient Boosting Decision Tree (GBDT) are employed for feature selection, ensuring that the variables input into the model possess the strongest predictive capabilities. In the prediction phase, Bidirectional Gated Recurrent Unit (BIGRU) \cite{BIGRU42} is utilized for point prediction, leveraging bidirectional layers to capture the complex dynamic patterns within the time series. Following this, Gaussian Process Regression (GPR) \cite{GPR42} is applied for interval prediction, providing confidence intervals for the forecast results \cite{carbon42}.

Wang et al. proposed a carbon price interval prediction model based on Lower Upper Bound Estimation (LUBE), Asymmetric Multi-Objective Evolutionary Algorithm (MOEA), and Long Short-Term Memory networks \cite{lstm45}. First, the carbon price time series is decomposed using Complete Ensemble Empirical Mode Decomposition with Adaptive Noise to reduce noise and extract key features. Next, Symbol Transfer Entropy (STE) is employed for causal inference and feature selection to ensure the effectiveness of the input variables \cite{ste58}. Then, based on the prediction points generated by LSTM, the LUBE model uses MOEA to optimize parameters and generate asymmetric prediction intervals \cite{np17}.

Zhang et al. proposed an enhanced decomposition-ensemble model for both point and probabilistic carbon price forecasting. The model integrates Complete Ensemble Empirical Mode Decomposition with Adaptive Noise, Fuzzy Entropy (FE) \cite{fe67}, Variational Mode Decomposition \cite{vmd66}, Long Short-Term Memory networks, Gated Recurrent Unit networks, Cuckoo Search Algorithm (CS) \cite{cs68}, and Gaussian Process Regression. Initially, CEEMDAN is employed to decompose the carbon price series into Intrinsic Mode Functions and residuals, followed by FE to quantify the complexity of these IMFs. High-complexity IMFs undergo further decomposition using VMD. LSTM and GRU models then predict each decomposed component, with CS optimizing their weights. Finally, GPR is used to generate prediction intervals, offering probabilistic insights \cite{np13}. 

Xu et al. proposed a hybrid framework for real-time carbon price prediction, tackling data feature drift and noise challenges in time series decomposition. The model uses Real-Time Complete Ensemble Empirical Mode Decomposition (RT-EMD) to decompose time series into meaningful components and residuals, which are further processed with Fuzzy Information Granulation (FIG) to reduce complexity. Key features are selected using Mutual Information Association Features (MIAF), and predictions are made using a Regularized Extreme Learning Machine (RELM) for deterministic forecasting and Gaussian Process Regression for probabilistic intervals~\cite{xu2024novel}. 

Zheng et al. proposed a decomposition-ensemble model for carbon price point and interval forecasting, integrating Sequential Variational Mode Decomposition (SVMD), Sample Entropy (SE), Singular Spectrum Analysis (SSA), and machine learning models. Initially, SVMD is employed to decompose the carbon price time series into Intrinsic Mode Functions and residuals. Subsequently, SE is utilized to evaluate the complexity of these IMFs, classifying them into low-frequency and high-frequency components. During the point forecasting stage, low-frequency components are predicted using a CatBoost model optimized with SSA, while high-frequency components are predicted using a Kernel Extreme Learning Machine, also optimized with SSA. After completing the predictions for both low-frequency and high-frequency components, their results are combined through direct summation to generate the final point forecasts. Finally, Adaptive Bandwidth Kernel Density Estimation (ABKDE) is applied to the residuals to estimate the error distribution and generate probabilistic intervals, providing uncertainty quantification and robust forecasting results~\cite{zheng2024multifactor}.

Ji et al. proposed a novel probabilistic carbon price prediction framework by integrating mixed-frequency modeling with a Transformer-based quantile regression architecture, aiming to improve both point forecasting accuracy and uncertainty quantification of carbon prices \cite{ji2025novel}. First, the framework incorporates multi-source explanatory variables with heterogeneous sampling frequencies, including energy, financial, macroeconomic, and environmental factors, which are transformed into unified representations through a mixed-data sampling (MIDAS) approach. This mixed-frequency modeling strategy effectively preserves high-frequency information while maintaining consistency with low-frequency carbon price series. Subsequently, a Transformer network is employed to capture complex nonlinear temporal dependencies within the mixed-frequency features, leveraging its self-attention mechanism to model long-range interactions in the carbon price dynamics. To further characterize predictive uncertainty, quantile regression is embedded into the Transformer framework, enabling the direct estimation of carbon price distributions at different quartiles rather than relying solely on point predictions. Finally, kernel density estimation is applied to the quantile-based outputs to construct continuous probabilistic forecasts. Empirical results demonstrate that the proposed MIDAS--Transformer--quantile framework outperforms multiple benchmark models in terms of both predictive accuracy and interval reliability across several carbon markets \cite{ji2025novel}.

Overall, this group marks an important shift from accuracy-oriented point prediction to risk-aware carbon price forecasting. Compared with earlier studies that mainly evaluate models through RMSE, MAE, or MAPE, these models introduce interval and probabilistic indicators such as PICP, PINAW, CWC, and AIS, as later summarized in Table~\ref{tab:eval_matrix}. This change is important for FinTech applications because financial managers and carbon market participants need to evaluate not only the expected price but also the possible range and uncertainty of future carbon credit prices. At the same time, this group also shows that uncertainty modeling does not automatically solve all methodological risks. Interval reliability, probabilistic calibration, mixed-frequency alignment, and leakage control remain critical issues for real-world deployment.

\medskip
\noindent\textbf{Deployment-oriented advanced architectures with interpretability, robustness, and external information.}

The fourth group of studies reflects a more deployment-oriented stage in carbon price forecasting. Instead of focusing only on reducing point prediction errors, these studies further consider interpretability, robustness, external information, graph-based relationships, transformer-based architectures, and advanced short-term forecasting frameworks.

Deng et al. proposed a carbon price trend prediction model that integrates several advanced techniques. The model employs Complete Ensemble Empirical Mode Decomposition with Adaptive Noise to decompose carbon price time series into Intrinsic Mode Functions and residuals, allowing the extraction of significant features. Following this, a Two-Stage Feature Selection (TFS) method is used to identify the most relevant features from these IMFs, which are then fed into a Light Gradient Boosting Machine (LightGBM) model \cite{lightgbm60}. The hyperparameters of the LightGBM model are optimized using Bayesian Optimization Algorithm (BOA) to enhance its predictive performance \cite{boa61}. Finally, SHapley Additive exPlanations (SHAP) is applied to interpret the model’s predictions \cite{shap62}, providing insights into the contributions of various features to the predicted carbon price trends \cite{np16}.

Shi et al. proposed a deep learning framework combining Convolutional Neural Networks (CNN) and Long Short-Term Memory networks to predict carbon prices with high accuracy and robustness. The model first uses CNN layers to extract spatial features from the carbon price time series data, which are then passed to LSTM layers to capture temporal dependencies \cite{cnn63}. This hybrid CNN-LSTM model is designed to address the limitations of traditional models in capturing complex nonlinear patterns in carbon markets \cite{lstm64}. The study demonstrates that the CNN-LSTM model significantly outperforms single CNN or LSTM models in both prediction accuracy and robustness, particularly in handling anomalies and outliers in the data \cite{np15}.

Hong et al. proposed a hybrid carbon price forecasting framework that integrates a Deep Augmented Frequency Enhanced Decomposed Transformer (DA-FEDformer) with a Multimodel Optimization Piecewise Error Correction System \cite{da-fed69}. The DA-FEDformer model enhances frequency domain information extraction using a multi-layer perceptron integrated into the encoder-decoder layers, combined with an Improved Kernel Mean Square Error (IKMSE) loss function and the Lion Optimizer. This model generates initial carbon price predictions, which are then refined by the error correction system \cite{ICEEMDAN40}. The system employs Improved Complete Ensemble Empirical Mode Decomposition with Adaptive Noise (ICEEMDAN) to decompose the prediction error sequence and applies five different models to correct these errors \cite{np2}.

Mao and Yu proposed a hybrid carbon price forecasting model integrating CEEMDAN, VMD, Sample Entropy, LightGBM, and Grey Wolf Optimizer (GWO). The model employs CEEMDAN to decompose the carbon price series into Intrinsic Mode Functions and residuals, followed by VMD for further refinement of high-noise IMFs. SE is then used to classify the IMFs into high-frequency, medium-frequency, and low-frequency components, which are individually predicted using LightGBM models optimized by GWO. The predictions from all components are aggregated into the final point forecast, with SHAP applied to ensure model interpretability~\cite{mao2024hybrid}.

Zhang et al. proposed a text-based carbon price forecasting framework based on a multivariate temporal graph neural network (MtemGNN), aiming to enhance prediction accuracy by integrating unstructured textual information from online news headlines \cite{zhang2025text}. First, the framework incorporates historical carbon price data together with online news headline data as multi-source inputs. To extract effective textual features, a Dynamic Topic Model (DTM) is employed to capture time-varying thematic information from news across different periods, while sentiment information is extracted using SnowNLP and further optimized by introducing an accumulative decay factor to account for the temporal attenuation of news impact. Subsequently, a multivariate temporal graph neural network is constructed to automatically learn the complex interdependencies among carbon prices, news topics, and sentiment features in a data-driven manner without requiring prior knowledge of graph topology. By simultaneously modeling temporal autocorrelation and inter-feature relationships, the proposed MtemGNN framework demonstrates superior carbon price forecasting performance compared with multiple benchmark models across several regional carbon markets in China \cite{zhang2025text}.

Wang et al. proposed a carbon trading price prediction framework based on a dual decomposition strategy that integrates Complex Wavelet Transform (CWT) and Gaussian Process Regression (GPR), aiming to enhance prediction accuracy by effectively handling nonlinear and non-stationary characteristics of carbon price series \cite{wang2025carbon}. First, the carbon price time series is decomposed using CWT to extract multi-scale components, with Kolmogorov complexity employed to determine the optimal number of decomposition layers and Approximate Entropy used to reconstruct high-frequency components with significant randomness. Subsequently, the residual sequences obtained from the first decomposition are further decomposed through GPR, which separates trend and high-frequency kernel components by maximizing the log marginal likelihood, forming a second-stage decomposition process. After the twice decomposition, a Gated Recurrent Unit (GRU) model is utilized to predict each decomposed component, while the Crested Porcupine Optimization (CPO) algorithm is applied to optimize the model parameters. Finally, the predicted results of all components are aggregated to generate the final carbon trading price forecasts. Experimental results based on carbon trading data from Wuhan and Beijing demonstrate that the proposed CWT--GPR--GRU framework achieves superior predictive performance compared with several benchmark models, particularly in short-term forecasting horizons \cite{wang2025carbon}.

Overall, this group shows that recent carbon price forecasting research is moving beyond conventional decomposition-based accuracy improvement toward more deployment-oriented concerns. Interpretability methods such as SHAP help explain which features drive predicted price movements, while robustness evaluation, text-based information extraction, graph learning, and transformer-based architectures expand the practical scope of forecasting models. As later reflected in Table~\ref{tab:eval_matrix}, these studies also diversify the types of inputs and model structures used in carbon price prediction, including national market data, news headlines, trend labels, and advanced neural architectures. However, these advances also introduce new methodological challenges, including dependence on text preprocessing quality, model reproducibility, market-specific feature construction, and inconsistent validation designs across studies. Therefore, their value lies not only in improving forecasting performance, but also in showing how carbon price prediction models can be made more interpretable, information-rich, and closer to real FinTech deployment requirements.

The above four groups show that carbon price forecasting methods have evolved from basic decomposition-based point prediction toward optimized hybrid models, uncertainty-aware forecasting, and more deployment-oriented architectures. To consolidate the technical characteristics of the reviewed studies, Table~\ref{tab:price-prediction-methods} summarizes their main price prediction methods, optimization algorithms, data decomposition techniques, and use of multi-factor inputs.

\begin{footnotesize}
\setlength{\tabcolsep}{4pt}
\renewcommand{\arraystretch}{1.65}

\begin{longtable}{@{\extracolsep{\fill}}
  >{\raggedright\arraybackslash}p{0.08\textwidth}  
  >{\raggedright\arraybackslash}p{0.26\textwidth}  
  >{\raggedright\arraybackslash}p{0.24\textwidth}  
  >{\raggedright\arraybackslash}p{0.32\textwidth}  
  >{\centering\arraybackslash}p{0.10\textwidth}    
@{}}
\caption{Summary of Price Prediction Methods and Techniques}
\label{tab:price-prediction-methods}\\
\toprule
Ref. & Price prediction method & Optimization method & Data decomposition technique & Multi-factor \\
\midrule
\endfirsthead

\toprule
Ref. & Price prediction method & Optimization method & Data decomposition technique & Multi-factor \\
\midrule
\endhead

\midrule
\multicolumn{5}{r}{\textit{(Continued on next page)}}\\
\midrule
\endfoot

\bottomrule
\endlastfoot

\cite{carbon38} & LS-SVM & Improved Particle Swarm Optimization (IPSO) & EMD & Y \\
\cite{carbon39} & LSTM, GRU, MLP, BPNN & Genetic Algorithm (GA) & CEEMDAN+Permutation Entropy+Variational Mode Decomposition & N \\
\cite{carbon40} & KELM & Multi-Objective Chaotic Sine Cosine Algorithm (MOCSCA) & ICEEMDAN & Y \\
\cite{carbon41} & SVR (high-frequency subsequences) + MLP (low-frequency subsequences) & - & ICEEMDAN+Discrete Wavelet Transform & Y \\
\cite{carbon42} & BIGRU point prediction + GPR interval prediction & - & CEEMDAN+Wavelet Transform & Y \\
\cite{carbon43} & GRU & - & Empirical Wavelet Transform & N \\
\cite{carbon44} & Extreme Learning Machine (ELM) & Cooperative Kidney Algorithm (CKA) & Mean Optimization Empirical Mode Decomposition (MOEMD) & Y \\
\cite{carbon45} & LSTM & - & MEEMD & Y \\
\cite{carbon47} & XGBoost, RF, KNEA, SVM, RBFNN & GWO, PSO, SA, FFOA & CEEMDAN & N \\
\cite{carbon48} & Weighted Regularized Extreme Learning Machine, WRELM & Multi-Objective Grasshopper Optimization Algorithm, MOGOA & ICEEMDAN & Y \\
\cite{carbon49} & LSTM & IWOA & MEEMD & N \\
\cite{carbon51} & VGM (long-term trend), ELM (short-term trend) & PSO & Hodrick-Prescott, HP filter & Y \\
\cite{np17} & LSTM, Lower Upper Bound Estimation (LUBE) & Asymmetric Multi-Objective Evolutionary Algorithm, MOEA & CEEMDAN & Y \\
\cite{np16} & Light Gradient Boosting Machine (LightGBM) & Bayesian Optimization Algorithm, BOA & CEEMDAN, Two-Stage Feature Selection, TFS & Y \\
\cite{np15} & LSTM & - & Convolutional Neural Network & N \\
\cite{np14} & LSTM & - & Multivariate Variational Mode Decomposition, MVMD & Y \\
\cite{np13} & LSTM, GRU & Cuckoo Search Algorithm, CS & CEEMDAN, Fuzzy Entropy, FE, Variational Mode Decomposition & Y \\
\cite{np2} & Deep Augmented Frequency Enhanced Decomposed Transformer (DA-FEDformer), Multimodel Optimization Piecewise Error Correction System & - & ICEEMDAN & N \\
\cite{xu2024novel} & RELM (global and local sequence prediction), GPR (interval prediction) & - & Real-Time EMD, FIG & N \\
\cite{zheng2024multifactor} & CatBoost (low-frequency), KELM (high-frequency) & Singular Spectrum Analysis (SSA) & Sequential Variational Mode Decomposition, Sample Entropy & Y \\
\cite{mao2024hybrid} & LightGBM & Grey Wolf Optimizer & CEEMDAN, VMD, Sample Entropy & Y \\
\cite{zhang2025text} & Multivariate Temporal Graph Neural Network (MtemGNN) & - & Dynamic Topic Model, sentiment feature construction & Y \\
\cite{ji2025novel} & Transformer-based quantile regression & - & Mixed-Data Sampling (MIDAS) & Y \\
\cite{wang2025carbon} & GRU & Crested Porcupine Optimization (CPO) & Complex Wavelet Transform, Gaussian Process twice decomposition & N \\

\end{longtable}
\end{footnotesize}

As systematically summarized in Table~\ref{tab:price-prediction-methods}, recent research on carbon credit price prediction shows a clear shift toward hybrid forecasting architectures. Most reviewed studies no longer rely on a single forecasting model, but combine data decomposition techniques with machine learning or deep learning predictors. Decomposition methods such as EMD, CEEMDAN, ICEEMDAN, VMD, wavelet transforms, and HP filtering are widely used to reduce multi-scale noise and extract more stable sub-series from nonlinear and non-stationary carbon price data. These decomposition layers are often further combined with optimization algorithms, such as IPSO, GA, MOGOA, GWO, BOA, PSO, and CPO, to tune model parameters and improve reported forecasting performance.

The table also shows that multi-factor modeling has become a common feature in carbon price prediction studies. In addition to historical carbon prices, many models incorporate external variables, including energy prices, macroeconomic indicators, market-related variables, and textual information, to capture broader drivers of carbon price fluctuations. At the same time, the reviewed studies reveal a gradual expansion from deterministic point prediction to more informative forecasting outputs, including interval and probabilistic prediction through methods such as GPR, LUBE, ABKDE, and quantile regression. Overall, Table~\ref{tab:price-prediction-methods} provides a technical summary of how the reviewed models combine predictors, decomposition methods, optimization strategies, and multi-source inputs.

\subsection{Critical Synthesis and Methodological Pitfalls}

To provide a consistent evaluation frame across all forecasting studies reviewed in Section~\ref{chp:algorithmsCC}, Table~\ref{tab:eval_matrix} systematically compares the reviewed models in terms of target market, data frequency, forecasting horizon, validation design, baseline models, evaluation metrics, statistical testing, and methodological limitations. Compared with a purely narrative review, this matrix-based synthesis makes the comparability of the reviewed models explicit and allows the forecasting literature to be assessed not only by reported error reduction, but also by the credibility and practical transferability of the experimental design.

\begin{footnotesize}
\setlength{\tabcolsep}{3pt}
\renewcommand{\arraystretch}{1.35}

\begin{longtable}{@{\extracolsep{\fill}}
  >{\raggedright\arraybackslash}p{0.08\textwidth}
  >{\raggedright\arraybackslash}p{0.16\textwidth}
  >{\raggedright\arraybackslash}p{0.20\textwidth}
  >{\raggedright\arraybackslash}p{0.18\textwidth}
  >{\raggedright\arraybackslash}p{0.18\textwidth}
  >{\raggedright\arraybackslash}p{0.16\textwidth}
@{}}
\caption{Consistent Evaluation Frame for Reviewed Carbon Price Forecasting Studies}
\label{tab:eval_matrix}\\
\toprule
\textbf{Ref.} & \textbf{Market \& Freq.} & \textbf{Horizon \& Validation Design} & \textbf{Baselines} & \textbf{Metrics \& Stat. Tests} & \textbf{Addressed Pitfalls / Limitations} \\
\midrule
\endfirsthead

\toprule
\textbf{Ref.} & \textbf{Market \& Freq.} & \textbf{Horizon \& Validation Design} & \textbf{Baselines} & \textbf{Metrics \& Stat. Tests} & \textbf{Addressed Pitfalls / Limitations} \\
\midrule
\endhead

\midrule
\multicolumn{6}{r}{\textit{(Continued on next page)}}\\
\midrule
\endfoot

\bottomrule
\endlastfoot

Sun \& Wang \cite{carbon38} &
Hubei, Shenzhen, Beijing \newline (Daily) &
1-step ahead; \newline Static train-test split &
ELM, BPNN, LSSVM, IPSO-LSSVM, EMD/WT variants &
MAPE, RMSE, MAE, R$^2$ \newline \textit{Stat. Test: Not reported} &
Uses factor analysis and EMD to reduce input complexity; static split and full-sample decomposition may still create leakage risk. \\

Nadirgil \cite{carbon39} &
EU ETS / EUA \newline (Daily) &
1-step ahead; \newline Static train-test split &
Single and hybrid LSTM, GRU, MLP, BPNN variants with CEEMDAN/VMD and GA &
R$^2$, RMSE, MAE, MAPE \newline \textit{Stat. Test: Not reported} &
Compares 48 hybrid models; strong accuracy but high model-selection and tuning complexity. \\

Hao \& Tian \cite{carbon40} &
Hubei, Shenzhen \newline (Daily) &
1-step and multi-step; \newline Static train-test split; repeated runs &
KELM, ELM, ANN and multi-objective optimized variants &
MAE, RMSE, MAPE, IA, U1 \newline \textit{Stat. Test: DM test} &
Adds multi-factor inputs, feature selection and MOCSCA optimization; complex pipeline increases tuning burden. \\

Li \& Liu \cite{carbon41} &
Hubei, Guangdong, Shenzhen \newline (Daily) &
1-step ahead; \newline Static train-test split &
SVR, MLP and decomposition / feature-screening variants &
MAE, RMSE, MAPE, SMAPE, TIC \newline \textit{Stat. Test: Not reported} &
Uses secondary decomposition and three-stage feature screening; decomposition-before-split risk should be considered. \\

Wang et al. \cite{carbon42} &
Shanghai, Hubei, Guangdong \newline (Daily) &
Point and interval prediction; \newline Static train-validation-test split &
PCA/RF/GBDT feature screening variants; BIGRU and GPR variants &
RMSE, MAE, MAPE, CP, MWP \newline \textit{Stat. Test: Not reported} &
Combines point and interval prediction; interval module improves uncertainty reporting but adds model complexity. \\

Liu \& Shen \cite{carbon43} &
EU ETS / EUA futures \newline (Daily) &
1-step ahead; \newline Static split across contract samples &
ARIMA, BPNN, GRU, EWT-BPNN &
RMSE, MAE, Dstat \newline \textit{Stat. Test: Not reported} &
EWT stabilizes non-stationary series; limited external factors and static validation. \\

Huang \& He \cite{carbon44} &
8 China pilots \newline (Daily) &
1-step ahead; \newline Static train-test split &
ELM, KA-ELM, CKA-ELM and structured/unstructured input variants &
RMSE, MAE, MAPE, R$^2$ \newline \textit{Stat. Test: Not reported} &
Introduces Baidu-index unstructured factors; interpretability of search-index selection remains limited. \\

Min et al. \cite{carbon45} &
Hubei, Shanghai, Beijing, Guangdong, Shenzhen \newline (Daily) &
1-step ahead; \newline Static train-test split &
LSTM, single-factor MEEMD-LSTM and other benchmark models &
RMSE, MAE, R$^2$ \newline \textit{Stat. Test: Not reported} &
Uses multi-factor MEEMD-LSTM and production rules; still relies on static validation. \\

Lu et al. \cite{carbon47} &
8 China pilots \newline (Daily) &
1-step ahead; \newline Static train-test split &
CEEMDAN-GWO-KNEA, CEEMDAN-PSO-SVM, CEEMDAN-SA-FFOA-SVM, CEEMDAN-RBFNN, CEEMDAN-RF, CEEMDAN-XGBoost &
MAE, RMSE, MAPE, RMSPE, U1, U2, Accuracy, SDE \newline \textit{Stat. Test: Not reported} &
Broad market coverage; reports stability, but some SVM/tree models show overfitting in specific markets. \\

Hao et al. \cite{carbon48} &
EU ETS and China market \newline (Daily) &
1-step ahead; \newline Static train-test split &
WRELM, ELM, ANN, ENN and optimized variants &
MAE, RMSE, MAPE, IA, U1, DM test \newline \textit{Stat. Test: DM test} &
Balances accuracy and stability via MOGOA; limited to two markets. \\

Yang et al. \cite{carbon49} &
Beijing, Fujian, Shanghai \newline (Daily) &
1-step ahead; \newline Static train-test split &
11 benchmark models, including LSTM and decomposition / optimization variants &
MAE, MAPE, RMSE, R$^2$ \newline \textit{Stat. Test: Not reported} &
MEEMD and IWOA improve LSTM forecasting; many hyperparameters increase tuning risk. \\

Zhao et al. \cite{carbon51} &
EU ETS / EUA futures \newline (Monthly) &
Next-step forecasting; \newline Static train-test split &
GM, ELM, VGM-ELM and 10 benchmark models &
MAE, MAPE, RMSE, Success ratio, IA, U1 \newline \textit{Stat. Test: DM test} &
Separates long-term trend and short-term fluctuation; monthly frequency supports planning but limits short-term trading use. \\

Wang et al. \cite{np17} &
9 China markets \newline (Daily) &
Short-term interval prediction; \newline Static train-test split &
LUBE variants; NSGA-II-LSTM, BP, PSO-LSTM and ensemble methods &
ACE, AIW, PICP/PINAW-related interval metrics \newline \textit{Stat. Test: HW test} &
Addresses asymmetric interval uncertainty and causal feature selection; interval quality depends on multi-objective search. \\

Deng et al. \cite{np16} &
Hubei \newline (Daily) &
Multi-step trend prediction; \newline Static train-test split &
LightGBM, BOA-LightGBM, CEEMDAN and TFS ablation variants &
AUC, Accuracy, Precision, Recall, F1 \newline \textit{Stat. Test: Not reported} &
Focuses on directional trend and SHAP interpretability; predicts direction rather than exact price level. \\

Shi et al. \cite{np15} &
Shenzhen \newline (Daily) &
1-step ahead; \newline Static train-test split &
CNN, LSTM and selected advanced models for comparison &
MSE, MAE, R$^2$, Z-score \newline \textit{Stat. Test: Not reported} &
Explicitly evaluates robustness with Z-score; single-market design limits generalizability. \\

Zhang et al. \cite{np14} &
EU ETS datasets \newline (Daily) &
Multi-step ahead; \newline Static train-test split &
ET, LSTM, MVMD and advanced hybrid baselines &
MAE, RMSE, MAPE, R$^2$ and stability indicators \newline \textit{Stat. Test: Not reported} &
Uses multivariate decomposition for multi-step forecasting; decomposition-based design may face leakage if not implemented chronologically. \\

Zhang \& Wang \cite{np13} &
Guangdong, Hubei, Tianjin, Shanghai, Beijing \newline (Daily) &
Point and interval prediction; \newline 60/20/20 train-validation-test split &
EMD-PSO-LSSVR, SD-LSTM, CEEMDAN-VMD-SE-LSTM and single-model variants &
R$^2$, MSE, MAE, MAPE, PICP, PIAW/PINAW \newline \textit{Stat. Test: Not reported} &
Combines deterministic and probabilistic forecasting; complex stacking may increase overfitting and tuning risk. \\

Hong et al. \cite{np2} &
Beijing, Guangzhou, Hubei \newline (Daily) &
1-step ahead; \newline 60/20/20 train-validation-test split &
ARIMA, LSTM, ELMAN, Informer, DA-FEDformer &
RMSE, MAE, MAPE, R$^2$ \newline \textit{Stat. Test: Not reported} &
Uses error correction to reduce residual bias; post-processing improves accuracy but increases pipeline complexity. \\

Xu et al. \cite{xu2024novel} &
Guangdong, Hubei \newline (Daily) &
1-, 3-, and 5-step; \newline Real-time decomposition with train-validation-test split &
One-time decomposition, real-time EMD, CEEMDAN/VMD/WPD-based variants &
MAE, RMSE, NRMSE, PICP, PINAW, CWC, AIS \newline \textit{Stat. Test: Not reported} &
Directly addresses decomposition leakage and feature drift via real-time decomposition. \\

Zheng et al. \cite{zheng2024multifactor} &
Hubei \newline (Daily) &
1-step point and interval prediction; \newline Static train-test split &
VMD, SVMD, CatBoost, KELM, KDE and combination variants &
MAE, MAPE, RMSE, R$^2$, PICP, PINAW \newline \textit{Stat. Test: Not reported} &
Adds SHAP and adaptive bandwidth KDE; single-market and single-step design limit broader validation. \\

Mao \& Yu \cite{mao2024hybrid} &
China national carbon market \newline (Daily) &
1-step ahead; \newline Static train-test split &
LightGBM and decomposition / reconstruction / feature-selection variants &
MAE, RMSE, MAPE, R$^2$ and interpretability analysis \newline \textit{Stat. Test: Not reported} &
Focuses on national market and interpretability; national market history is short, increasing small-sample risk. \\

Zhang et al. \cite{zhang2025text} &
Shenzhen, Guangzhou, Fujian \newline (Daily + news headlines) &
1-step ahead; \newline Static train-test split &
XGBoost, LightGBM, GRU, LSTM, Transformer, Informer, VGN, GNN &
MAE, RMSE, R$^2$ \newline \textit{Stat. Test: Not reported} &
Uses DTM/SnowNLP and MtemGNN to model text-driven factors; quality depends on news-source coverage and NLP extraction. \\

Ji et al. \cite{ji2025novel} &
Shanghai, Hubei \newline (Weekly target + daily factors) &
Point, interval and probabilistic prediction; \newline Mixed-frequency train-test design &
QRNN, QRLSTM, QRGRU, QRTransformer and MIDAS variants &
MAE, RMSE, MAPE, TIC, R$^2$, PICP, PINAW, CWC, AIS \newline \textit{Stat. Test: Not reported} &
Addresses uncertainty and mixed-frequency information; avoids decomposition leakage but requires careful frequency alignment. \\

Wang et al. \cite{wang2025carbon} &
Wuhan, Beijing \newline (Daily) &
1-step ahead; \newline Static train-test split &
GRU, GPR and decomposition / optimization variants &
MAE, RMSE, MAPE, R$^2$ \newline \textit{Stat. Test: Not reported} &
Uses CWT-GPR twice decomposition and CPO-GRU; conference-length study gives limited validation detail. \\

\end{longtable}
\end{footnotesize}

Based on the complete evaluation matrix in Table~\ref{tab:eval_matrix}, several cross-study patterns can be identified. Although most reviewed models report improved forecasting accuracy relative to their own baselines, the strength of evidence varies substantially across studies because of differences in market coverage, forecasting horizon, validation strategy, baseline selection, and the use or absence of statistical testing. Therefore, the following synthesis focuses not only on prediction accuracy, but also on whether the evaluation design is sufficient to support real-world FinTech deployment.

While the proliferation of decomposition-ensemble models has reduced commonly reported prediction error metrics, such as RMSE, MAE, and MAPE, the full evaluation matrix also reveals several methodological pitfalls that may limit their real-world applicability in carbon-related financial services.

\textbf{Overfitting Risk from Multi-Stage Complexity:} A clear trend across the reviewed studies is the increasing use of multi-stage hybrid pipelines, typically combining decomposition methods such as EMD, CEEMDAN, ICEEMDAN, VMD, or MVMD with feature selection, heuristic optimization, and deep learning predictors. These architectures can improve the fit to historical carbon price series by separating short-term noise from lower-frequency movements. However, they also substantially increase model capacity and the number of design choices. When such highly parameterized models are trained on relatively short carbon market datasets, especially in emerging or newly established markets, they may capture market-specific noise rather than generalizable price dynamics. This issue is particularly important because many studies still rely on static train-test splits and do not report formal statistical tests. Consequently, lower reported error values should not be interpreted as sufficient evidence of robust out-of-sample performance.

\textbf{Data Leakage via Decomposition and Tuning:} Another recurring concern is the risk of data leakage in time-series preprocessing and model selection. First, \textit{decomposition leakage} may occur when methods such as CEEMDAN, ICEEMDAN, VMD, or wavelet-based decomposition are applied to the full historical series before the training and testing periods are separated. In this case, information from future observations can indirectly affect the decomposition components used during model training. Second, \textit{tuning leakage} may occur when hyperparameter optimization or ensemble-weight selection is repeatedly adjusted according to final test-set performance. This effectively turns the test set into a validation set and may inflate reported forecasting accuracy. These leakage risks are not always explicitly discussed in the reviewed studies, even though they are highly relevant for time-series deployment in FinTech environments.

\textbf{Vulnerability to Regime Shifts:} Carbon markets are not purely statistical systems; they are heavily shaped by policy interventions, allowance allocation rules, compliance cycles, energy-market shocks, and transitions between trading phases. These factors can generate abrupt regime shifts in price dynamics. However, Table~\ref{tab:eval_matrix} shows that many studies still evaluate their models using static historical splits rather than rolling, walk-forward, or regime-aware validation designs. A model optimized under a relatively stable or low-volatility period may therefore perform poorly when deployed during policy tightening, quota adjustment, or sudden market stress. This limitation is especially relevant for carbon credit pricing because the target variable is directly connected to regulatory design rather than only to endogenous market behavior.

Recent studies have begun to address some of these weaknesses. Xu et al.~\cite{xu2024novel}, for example, explicitly identify the feature-drift and leakage problems caused by conventional one-time decomposition and propose a real-time decomposition framework to reduce future-data contamination. Ji et al.~\cite{ji2025novel} move the literature beyond deterministic point forecasting by combining mixed-frequency modeling, Transformer-based quantile regression, and probabilistic prediction. Rather than only minimizing point-error metrics, this type of uncertainty-aware design provides information about the possible distribution of future carbon prices, which is more aligned with risk-sensitive financial decision-making.

From a FinTech perspective, algorithm selection should be linked to the intended corporate or financial use case. For short-term trading, dynamic hedging, and operational risk monitoring, models require strong out-of-sample validation, careful leakage control, and robustness to market noise. In this context, real-time decomposition frameworks and short-horizon hybrid models, such as the RT-EMD-based model in Xu et al.~\cite{xu2024novel} or the CWT-GPR-GRU framework in Wang et al.~\cite{wang2025carbon}, are more relevant than models evaluated only through static historical splits. By contrast, for long-term corporate budget planning, carbon asset valuation, and discounted cash flow analysis, deterministic point forecasts are often insufficient. Probabilistic and interval forecasting models, such as LUBE-based interval prediction~\cite{np17} and quantile-regression-based probabilistic forecasting~\cite{ji2025novel}, provide more useful information because they allow financial managers to estimate carbon risk premiums and prepare for uncertainty under changing policy regimes.

\section{Prediction of Enterprise Carbon Emissions}
\label{chp:algorithmCE}
Carbon emission prediction algorithms are of paramount importance in the global effort to combat climate change. To ensure a rigorous analytical scope, it is essential to distinguish between the two primary components of an organization's carbon footprint: \textit{corporate operational carbon} and \textit{infrastructure embodied carbon}. In this review, \textit{enterprise carbon emissions} primarily refer to operational emissions generated during a company's daily activities, typically categorized into Scope 1 (direct emissions), Scope 2 (indirect emissions from purchased energy), and Scope 3 (all other indirect emissions in the value chain). Sections 6.1 and 6.2 will focus on predictive methodologies for these operational scopes. Beyond these immediate activities, carbon emissions associated with the physical assets owned or utilized by enterprises, specifically the embodied carbon in buildings and facilities, represent a significant yet distinct category of responsibility. This latter component, which encompasses emissions from the entire life cycle of construction materials, will be addressed separately in Section 6.3 to provide a comprehensive view of corporate environmental impact.

\subsection{\textbf{Empirical Models}}

In the context of predicting corporate-level carbon emissions, the study by Goldhammer et al. presents an innovative approach that estimates corporate carbon footprints (CCFs) using externally available data. They employ a regression analysis model that primarily utilizes five key variables as input data: firm size (measured by turnover or number of employees), vertical integration, capital intensity, production centrality, and the carbon intensity of the national energy mix. The model was trained on actual CCF data from 93 European organizations across the chemical, construction engineering, and industrial machinery sectors. The study demonstrates that this method can accurately estimate corporate carbon emissions without relying on internal data, making it particularly suitable for scenarios requiring rapid, external assessments \cite{carbon62}.

\cite{carbon64} explores the application of two primary technical methods. The first is based on the Economic Input-Output Life Cycle Assessment (EIO-LCA) model, which utilizes macroeconomic data, such as industry GDP and carbon emission intensity, to estimate corporate carbon emissions. The second method is the estimation model developed by MSCI ESG Research, which combines historical company disclosures with industry-specific intensity to predict carbon emissions for years without disclosed data. The research used the 2013 carbon emission data from 277 organizations in the MSCI ACWI IMI index as training data and evaluated the models' prediction accuracy by comparing them with the organizations' actual disclosed data. The results indicate that the MSCI ESG Research model outperforms the EIO-LCA model in estimation accuracy, particularly in handling company-specific carbon emission intensities \cite{carbon64}.

The study by Griffin et al. provides an effective technical approach for estimating corporate greenhouse gas (GHG) emissions using external data. This research employs the Ohlson valuation model \cite{ohlson65} and combines it with reported GHG emission data to develop a GHG emission estimation model through regression analysis. This model utilizes a range of company-specific variables as input data, including revenue, capital expenditures, and asset composition, to predict the emissions of organizations that have not disclosed their GHG data \cite{carbon63}.

\subsection{\textbf{Machine Learning Methods}}

Zhou et al. developed a carbon risk prediction model based on Support Vector Machines to accurately predict carbon emissions for heavily polluting industrial enterprises in China. The study first established a carbon risk assessment system comprising 45 initial indicators. After performing principal component analysis and independent analysis, 18 core indicators were selected. The SVM algorithm was used for model training, and the penalty parameter C and Gaussian kernel parameter gamma were optimized through grid search and cross-validation to handle the nonlinear and high-dimensional characteristics of the data \cite{ep2}.

Quyen et al. developed a novel approach for predicting corporate-level carbon emissions, specifically targeting organizations that do not publicly disclose their emissions data. The framework proposed a Meta-Elastic Net learning model to aggregate predictions from multiple base models, including Ordinary Least Squares (OLS), Elastic Net, Neural Networks, K-Nearest Neighbors (KNN), Random Forest, and Extreme Gradient Boosting. The input training data comprised global carbon emissions data from 2,289 organizations between 2005 and 2017, totaling 13,435 annual observations. The data included Scope 1, Scope 2, and Scope 3 emissions, as well as various corporate factors such as operational scale, business models, technological advancements, and energy consumption. The model's performance was validated through extensive cross-validation and robustness tests, demonstrating significant improvements in prediction accuracy compared to traditional regression models \cite{carbon65}.

Li et al. proposed an advanced carbon emission prediction model tailored for enterprise-level applications, integrating multi-dimensional feature extraction with an optimized Radial Basis Function (RBF) neural network. The model utilizes Principal Component Analysis \cite{pca} combined with kernel functions to extract essential features from a diverse set of input data, which includes production metrics, fuel consumption, electricity and heat energy purchases, and other relevant economic indicators. This approach effectively reduces data complexity while preserving key predictive information. The extracted features are then input into an RBF neural network, whose parameters are finely tuned using a hybrid optimization method that combines the Drosophila Optimization Algorithm and Differential Evolution. This dual optimization strategy significantly enhances the accuracy and robustness of the emission predictions \cite{ep3}.

Chen et al. proposed an efficient Fast Transfer-Based Adaptive Recurrent Neural Network (SEER) model aimed at addressing the high computational costs associated with real-time predictions in existing deep learning models. The SEER model enhances prediction accuracy and training efficiency by integrating data preprocessing, feature extraction, attention mechanisms, transfer learning, and adaptive learning rate mechanisms. Using carbon emission and air quality data from Shanghai and Beijing as inputs, the model effectively identifies key data points and optimizes generalization capabilities, demonstrating its accuracy and applicability in corporate carbon emission predictions \cite{ep7}.

Tang et al. proposed CarbonNet, a novel approach for enterprise-level carbon emission prediction, aimed at overcoming the limitations of traditional macro-level forecasting models. The paper first constructs a large-scale dataset encompassing 3346 organizations over 31 years, integrating carbon emission data with corporate financial and operational metrics. To enhance prediction accuracy, the authors employed Factor Analysis for feature extraction, reducing the impact of irrelevant variables, thereby building a more generalizable carbon emission prediction model. Subsequently, the paper utilized the XGBoost algorithm for large-scale data mining, iteratively optimizing prediction errors \cite{ep1}.

Shou developed a corporate carbon performance evaluation framework using Support Vector Machine to predict carbon emissions at the enterprise level. The framework, named Enterprise Carbon Performance Index System (ECPIS), integrates SVM with real-time data collection and monitoring to assess and forecast carbon emissions, enabling organizations to make data-driven decisions to reduce their carbon footprint. The input training data for the SVM model includes historical carbon emissions data, energy consumption metrics, operational data such as production rates and operating costs, as well as environmental data from air quality sensors and greenhouse gas analyzers. The system demonstrates a high accuracy in evaluating corporate carbon performance, outperforming other models like K-Nearest Neighbors, Random Forest, Decision Trees (DT), Naive Bayes (NB), and Logistic Regression \cite{ep5}.

Zhao et al. proposed an innovative approach for predicting carbon emissions at the enterprise level, specifically targeting coal combustion processes in power generation organizations. This model utilizes Rough Set Theory to preprocess and extract key features from incomplete and imprecise carbon emission data, thereby enhancing the reliability of the data for subsequent modeling. The model then integrates a Grey System GM(1,1) model with a Support Vector Machine to predict carbon emissions. The Grey model generates initial predictions based on historical data, which are further refined by the SVM to account for nonlinear relationships and improve accuracy. The input data for the model includes coal consumption rates, corresponding carbon emissions, and other related variables from a large coal-fired power company, covering multiple plants. The combination of the Grey model and SVM results in a hybrid prediction framework that significantly reduces prediction errors compared to traditional methods \cite{ep4}.

Shi et al. developed a carbon emission prediction model tailored for thermal power enterprises, utilizing data from 17 power plants in Gansu Province, China. The study introduced a hybrid approach that combines Factor Analysis (FA) with multiple machine learning models, including Support Vector Regression and a Genetic Algorithm-optimized Backpropagation Neural Network (GA-BP). The model inputs include critical variables such as energy consumption, electricity output, fuel type, and carbon content per unit of heat, derived from the 2021 greenhouse gas emissions reports and carbon verification documents of the power plants. Factor Analysis was employed to reduce data dimensionality by extracting key factors—such as energy consumption and output, energy quality, and energy efficiency—which were then used as inputs for the prediction models. Among the models evaluated, the GA-BP neural network demonstrated superior performance in predicting CO2 emissions, showcasing lower error rates and higher accuracy \cite{ep6}.

Pan et al. introduced a real-time estimation method for the cement industry, utilizing a CNN-BLSTM (Convolutional Neural Network–Bidirectional Long Short-Term Memory) model. This method accurately estimates both direct and indirect carbon emissions by integrating CNN for feature extraction from material and energy consumption data, and BLSTM \cite{Blstmep8} for capturing temporal dependencies. Additionally, the study employs a Marginal Carbon Emission Factor (MCEF) to assess indirect emissions. The model was trained using 30 days of operational data from a cement corporation, demonstrating superior accuracy compared to traditional methods \cite{ep8}.

\begin{figure}[ht]
\centering
\includegraphics[width=0.7\linewidth]{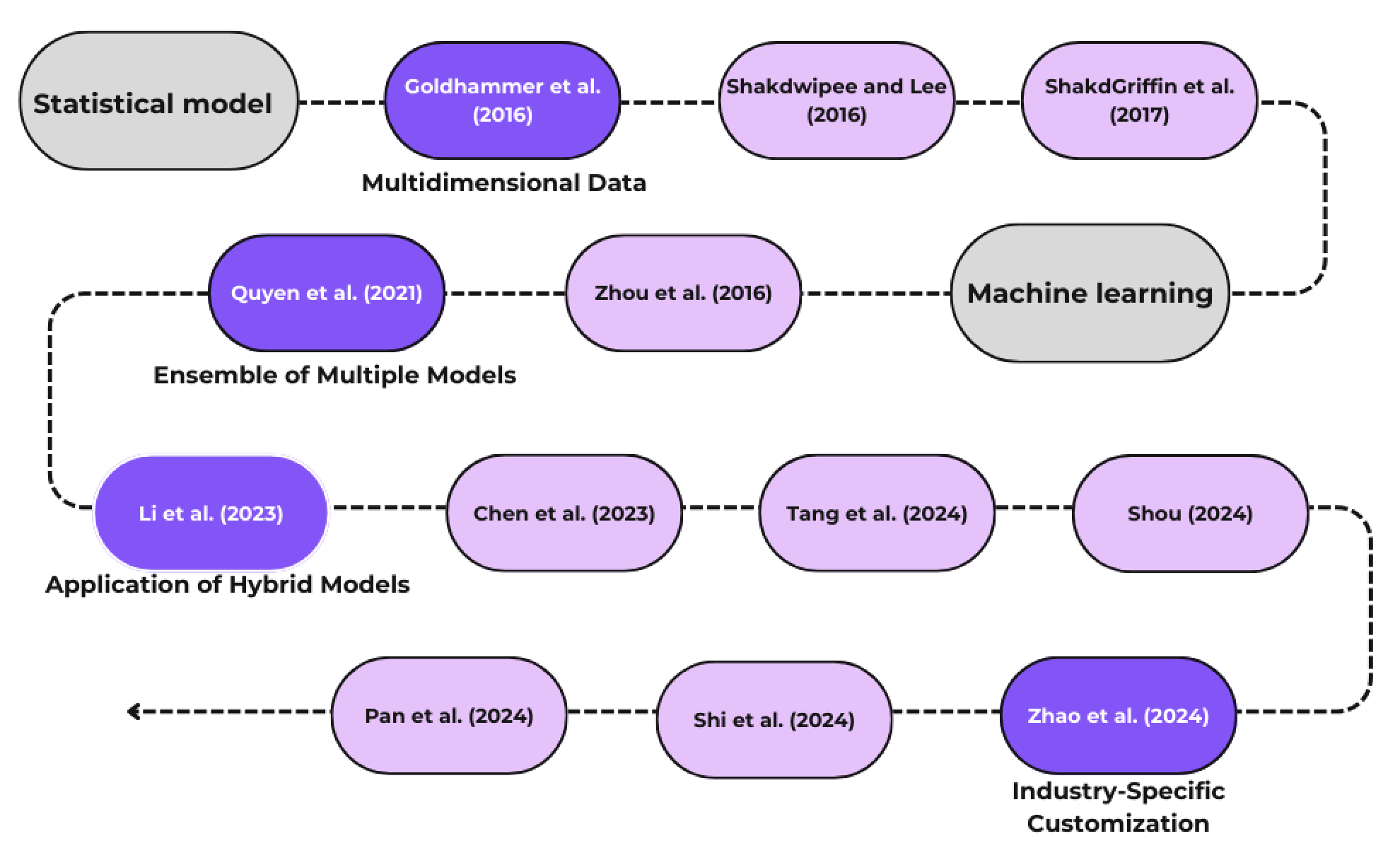}
\caption{Evolutionary Trends in Corporate Carbon Emission Prediction Algorithms}
\label{fig:Evolutionary_Trends}
\vspace{-.2in}
\end{figure}

After a thorough review of corporate carbon emission prediction algorithms, several conclusions could be drawn. The research in this field is increasingly moving towards more diverse data inputs, refined model processing, and industry-specific customization. By incorporating complex feature extraction techniques and multidimensional data inputs, researchers have enabled prediction models to more accurately capture the various factors influencing corporate carbon emissions. Moreover, the widespread application of machine learning and deep learning algorithms, along with the exploration of hybrid models and optimization algorithms, has significantly enhanced the accuracy and robustness of these prediction models. As the demand for real-time predictions grows, especially in high-emission industries, researchers have developed customized prediction models tailored to specific sectors, improving the effectiveness of these models, and Fig.~\ref{fig:Evolutionary_Trends} reflects the trends in the evolution of these technologies. These advancements not only assist organizations in better managing their carbon emissions but also provide investors and governments with more transparent and reliable carbon emission information, thereby supporting global efforts to achieve climate goals.

\subsection{Predictive Models for Embodied Carbon Emission}

In addition to carbon emissions generated during corporate operations, carbon emissions associated with buildings owned or utilized by enterprises should also be considered. Data indicates that emissions from building-related activities account for more than 39\% of global emissions, with embodied carbon contributing approximately 11\% \cite{unep20222022}. Therefore, accurately quantifying the embodied carbon within buildings is essential for comprehensively predicting carbon emissions and correctly attributing them to responsible enterprises.

Zhang et al. collected data from 850 building samples across China, covering structural characteristics and material usage. The dataset included 13 structural features such as building height, structural type, seismic design intensity, geographic location, delivery method, project cost, and material expenses. Additionally, 16 material-related features were recorded, including the consumption of steel, concrete, bricks, prefabricated components, doors, windows, and paints. The study applied a process-based Life Cycle Assessment methodology to calculate the Embodied Carbon Intensity during the material production stage. Outliers were identified and removed using the Isolation Forest algorithm, followed by feature selection through Spearman's correlation coefficient analysis to eliminate highly correlated variables. The researchers then employed 12 machine learning models for prediction, categorized as follows: linear models, Ridge Regression (RD), Bayesian Ridge Regression (BR), Elastic Net Regression (EN); single nonlinear models, K-Nearest Neighbors, Support Vector Regression; and ensemble models like Gradient Boosting, Extreme Gradient Boosting (XGB), Histogram-based Gradient Boosting, Random Forest, and Extremely Randomized Trees. The ET model demonstrated superior performance during the planning and design phase, with building height, material cost, and geographic region identified as the core predictive features. During the preliminary design stage, the XGB model achieved the highest performance, with material costs, rebar, and concrete usage emerging as the most influential factors \cite{embodiedcarbon1}.

Su et al. utilized data from 70 construction projects in the Yangtze River Delta region of China, capturing various building and construction-related characteristics. The data underwent preprocessing using the Quartile Method to eliminate outliers, followed by KNN-based imputation for missing values. Spearman's correlation analysis was employed to remove redundant features, ultimately yielding 15 project-level attributes (e.g., building height, location, structural type) and 12 construction-level attributes (e.g., on-site workforce, prefabricated concrete volume). SVR, ANN, and XGBoost models were applied to predict material consumption, which was then multiplied by corresponding Emission Factors to estimate total embodied carbon emissions. The model achieved a prediction error of less than 5.33\% \cite{embodiedcarbon3}.

Marsh et al. introduced a Monte Carlo Simulation–based approach to quantify uncertainties in embodied carbon (EC) calculations during the building product stage. The methodology begins by extracting material consumption data from construction inventories and carbon emission coefficients from established databases such as ICE and OneClickLCA. Outliers were removed using the Quartile Method, while Spearman's correlation analysis helped exclude redundant features. The study assumed that Embodied Carbon Coefficients (ECCs) follow a normal distribution, with missing values imputed based on the global mean coefficient of variation (CoV). Subsequently, 50,000 Monte Carlo simulations were conducted to generate a probabilistic distribution of EC predictions across different scenarios. The results included mean values, standard deviations, and confidence intervals to evaluate error margins resulting from data uncertainty. Sensitivity analysis revealed that high-variance materials, particularly rebar and steel, substantially influenced the prediction's dispersion. The findings emphasized that neglecting ECC uncertainty could lead to prediction deviations exceeding 10\%, underscoring the importance of incorporating uncertainty analysis during early design stages to enhance prediction accuracy and decision-making reliability \cite{embodiedcarbon4}.

For a comprehensive prediction of corporate carbon emissions, embodied carbon also plays a crucial role. Focusing solely on emissions generated during corporate operations while neglecting the carbon emissions embedded in buildings and infrastructure can lead to a systematic underestimation of the company's overall carbon footprint, ultimately compromising the scientific rigor and effectiveness of carbon management and policy-making. Currently, the primary method for predicting embodied carbon emissions in buildings is the Carbon Emission Factor approach, which involves identifying the individual components of building materials and assigning a corresponding carbon emission factor to each, thereby quantifying the carbon footprint of the building throughout its construction and life cycle stages. However, these carbon emission factors are influenced by various dynamic factors, such as production processes, material sources, and supply chain fluctuations, resulting in temporal variability. Consequently, researchers have increasingly adopted carbon emission uncertainty quantification methods, such as Monte Carlo simulations, to assess the potential prediction deviations, establishing this as a key research direction in the field. By systematically analyzing and predicting the embodied carbon characteristics in buildings, these research advancements not only support enterprises in formulating scientifically sound carbon management strategies but also provide critical insights that help market participants and regulatory bodies make informed decisions, ultimately contributing to the collective global response to climate change.
\section{Discussion and Conclusion}

\subsection{Comparative Synthesis and Contributions}

To clarify the unique positioning of this study and respond to recent developments in the field, we present a Scope Matrix in Table \ref{tab:scope_matrix} that compares this work with five prominent literature reviews published between 2022 and 2024. 

While the literature on carbon management is expanding, existing reviews typically operate within siloed disciplines. For instance, Wang et al. \cite{lr2} and Trouwloon et al. \cite{carbon75} provide excellent insights into the financial implications and governance of carbon risk/disclosure, yet they do not address the computational algorithms required to quantify these risks. Conversely, Alshatri \& Hussain \cite{carbon2} and Jin et al. \cite{lr1} focus on the technical drivers and forecasting models; however, Alshatri \& Hussain limit their scope to price drivers, and Jin et al.'s extensive review of 147 models focuses predominantly on national or city-level (macro) forecasting, offering limited utility for individual corporate decision-making.

\begin{table}[htbp]
\centering
\caption{Scope Matrix of existing literature reviews compared to this study}
\label{tab:scope_matrix}
\begin{footnotesize}
\setlength{\tabcolsep}{4pt}
\renewcommand{\arraystretch}{1.3}
\begin{tabular}{@{}
  p{0.18\textwidth}
  >{\centering\arraybackslash}p{0.15\textwidth}
  >{\centering\arraybackslash}p{0.12\textwidth}
  >{\centering\arraybackslash}p{0.15\textwidth}
  >{\centering\arraybackslash}p{0.16\textwidth}
  >{\centering\arraybackslash}p{0.16\textwidth}
  @{}}
\toprule
\textbf{Reference} & \textbf{Disclosure Impacts (Financial)} & \textbf{Price Drivers} & \textbf{Price Forecasting Methods} & \textbf{Enterprise Emissions Forecasting} & \textbf{FinTech / Corporate Decision Framing} \\
\midrule
Trouwloon et al. (2023) \cite{carbon75} & $\checkmark$ \newline (Focus on claims) & $\times$ & $\times$ & $\times$ & $\times$ \\
\addlinespace

Lokuge \& Anders (2022) \cite{carbon76} & $\times$ & $\times$ & $\times$ & $\times$ & $\times$ \newline (Agriculture focus) \\
\addlinespace

Alshatri \& Hussain (2023) \cite{carbon2} & $\times$ & $\checkmark$ & $\times$ \newline (Focus on drivers) & $\times$ & $\times$ \\
\addlinespace

Jin et al. (2024) \cite{lr1} & $\times$ & $\times$ & $\times$ & $\times$ \newline (Focus on macro/city) & $\times$ \\
\addlinespace

Wang et al. (2022) \cite{lr2} & $\checkmark$ \newline (Carbon risk) & $\times$ & $\times$ & $\times$ & $\times$ \\
\midrule

\textbf{This Study} & \textbf{$\checkmark$} & \textbf{$\checkmark$} & \textbf{$\checkmark$} & \textbf{$\checkmark$} \newline \textbf{(Enterprise/micro level)} & \textbf{$\checkmark$} \\
\bottomrule
\end{tabular}
\end{footnotesize}
\end{table}

To the best of our knowledge, this is the first systematic review that integrates the financial impacts of carbon disclosure, carbon price prediction algorithms, and enterprise-level (micro) carbon emission forecasting within a unified FinTech and corporate decision-making framework. This cross-disciplinary integration is critical for the "Responsible FinTech Era" as it bridges the gap between raw data science and actionable financial strategy. 

Specifically, this study surpasses existing literature in the following aspects:

\begin{enumerate}
    \item \textbf{A Multi-Dimensional Analytical Perspective:} Unlike reviews that focus solely on market performance or regulatory compliance, this study investigates how transparency in carbon management directly modulates corporate financing costs and investor trust, establishing a clear link between environmental responsibility and financial viability.

    \item \textbf{Focus on Micro-Level Predictive Technologies:} While existing reviews (e.g., Jin et al., 2024) often focus on macro-scale emissions, we systematically review technologies tailored for the enterprise level. This provides practical guidance for firms to assess their own carbon footprints and for investors to reduce information asymmetry.

    \item \textbf{Algorithmic Evaluation for Corporate Strategy:} Beyond cataloguing models, we evaluate the applicability of price and emission prediction algorithms specifically for optimizing corporate carbon credit procurement and budget planning.

    \item \textbf{Interdisciplinary Synthesis for Future Research:} By aligning data-driven precision with collective environmental welfare, this study provides a foundation for future quantitative research on how AI-enhanced carbon management can be discounted into corporate valuation.
\end{enumerate}

\subsection{Research Implications and Future Directions}

The existing literature has extensively studied the impact of carbon emissions and carbon information disclosure on corporate market value, financial performance, and loan interest rates. However, the role of carbon credits in corporate valuation, particularly in the financial and FinTech fields, remains an underexplored area. Consequently, we propose the following forward-looking research agenda, outlining how AI-driven predictive models can be operationalized into concrete financial metrics and integrated into established valuation frameworks.

Firstly, the application of carbon credits holds substantial importance for quantifying corporate market value. By predicting carbon emissions and carbon credit prices, organizations can accurately quantify future carbon liabilities on their balance sheets and project the impact of offsetting costs on operating margins (e.g., EBITDA). This forecasting not only helps organizations make more informed decisions regarding the timing and volume of carbon credit transactions but also allows for the development of automated FinTech procurement systems that optimize operational costs dynamically. In this process, organizations can effectively reduce carbon emission costs and voluntarily disclose carbon information, thereby enhancing market trust, lowering financing costs, and ultimately increasing market value. This research direction lays a strong foundation for future studies to quantify the long-term impact of carbon management on corporate market value.

Secondly, the impact of carbon credits on corporate valuation warrants further empirical investigation. By understanding the future costs incurred from purchasing carbon credits or the revenues generated from selling surplus allowances, financial analysts can explicitly integrate these variables into the Discounted Cash Flow (DCF) framework. Specifically, predicted carbon credit expenditures can be modeled as direct deductions from projected Free Cash Flows (FCF), while the algorithmic uncertainty of these predicted prices can be factored into the Weighted Average Cost of Capital (WACC) as a distinct carbon risk premium. This present value adjustment directly affects a company’s market value, helping managers gain a clearer understanding of how carbon emissions tangibly influence financial performance. This direction offers promising applications for building algorithmic trading and valuation tools within the FinTech sector.

Finally, dynamic carbon risk ratings represent a critical area of research. With the advancement of carbon emission prediction algorithms and price forecasting models, corporate carbon risks can be assessed with greater real-time accuracy. These high-frequency risk ratings can provide a basis for dynamic loan pricing models—such as automatically adjusting the basis points on corporate debt or green bonds based on real-time emission forecasts. By incorporating carbon emissions and carbon credit costs into asset pricing models (such as extended Fama-French multifactor models), predictions of corporate valuation will become more comprehensive and precise.

\subsection{Concluding Summary}

The study on carbon disclosure suggests that organizations should actively manage and accurately disclose their carbon emissions to mitigate the negative impact on market value and financial performance. By maintaining transparency, organizations can enhance market trust, reduce financing costs, and ultimately increase their market value.

Furthermore, research indicates that understanding the factors influencing carbon prices and accurately predicting carbon prices are crucial for corporate cost management. Accurate carbon price forecasts enable organizations to better plan the timing and volume of carbon credit transactions, thereby optimizing operational cost management.

Carbon emission predictions play a significant role in increasing corporate transparency. They not only help governments and financial investors gain a clearer understanding of actual corporate emissions but also provide a reliable basis for assessing environmental risks and future financial performance.

Finally, integrating carbon price predictions with carbon emission forecasts allows organizations to manage their carbon emission costs more comprehensively. The combination of these two techniques lays the foundation for future quantitative research, providing more precise tools for corporate valuation and market performance, helping organizations achieve sustainable development goals while enhancing market competitiveness. In this way, these advancements not only assist organizations in optimizing their carbon management strategies but also promote the development of the global carbon credit market, contributing to the achievement of sustainable development goals.

\section*{Funding Declaration}

The authors declare that no funding was received for this work.
\clearpage

\bibliography{carboncreditsurvey}

\end{document}